\documentclass[journal,draftcls,onecolumn,12pt,twoside]{IEEEtran}

\usepackage[utf8]{inputenc} 
\usepackage[T1]{fontenc}
\usepackage{url}
\usepackage{ifthen}
\usepackage{cite}
\usepackage{amsmath}%
\usepackage{wasysym}%
\usepackage{amssymb}

\usepackage{mdwmath}
\usepackage{blindtext}
\usepackage{eqparbox}
\usepackage{fixltx2e}
\usepackage{stfloats}

\usepackage[english]{babel}
\usepackage{lipsum}
\usepackage{tikz}
\usepackage{mathtools,amsfonts,amssymb,amsthm}
\usepackage{graphicx}

\usepackage{algorithm}
\usepackage{algpseudocode} 

\usepackage{geometry}
\newtheorem{theorem}{{Theorem}}

\newtheorem{lemma}[theorem]{{Lemma}}
\newtheorem{proposition}[theorem]{{Proposition}}

\newtheorem{definition}{{Definition}}

\theoremstyle{remark}
\newtheorem{remark}{{Remark}}

\newcommand{\cA}{{\cal A}} 
\newcommand{\cB}{{\cal B}}
\newcommand{\cC}{{\cal C}}

\newcommand{\cF}{{\cal F}}

\newcommand{\cR}{{\cal R}}
\newcommand{\cS}{{\cal S}} 
\newcommand{\cT}{{\cal T}}

\DeclareMathAlphabet{\mathbfsl}{OT1}{ppl}{b}{it} 

\newcommand{\bQ}{\mathbfsl{Q}} 
\newcommand{\bR}{\mathbfsl{R}}

\newcommand{\bU}{\mathbfsl{U}} 
\newcommand{\bV}{\mathbfsl{V}}
 
\newcommand{\bX}{\mathbfsl{X}}
 
\newcommand{\bZ}{\mathbfsl{Z}}

\newcommand{\bv}{\mathbfsl{v}}

\newcommand{\mathset}[1]{\left\{#1\right\}}
\newcommand{\abs}[1]{\left|#1\right|}

\def\QEDclosed{\mbox{\rule[0pt]{1.3ex}{1.3ex}}} 

\def\QED{\QEDclosed} 

\def\proof{\noindent\hspace{2em}{\itshape Proof: }}
\def\endproof{\hspace*{\fill}~\QED\par\endtrivlist\unskip}

\newcommand{\be}[1]{\begin{equation}\label{#1}}
\newcommand{\ee}{\end{equation}}

\renewcommand{\leq}{\leqslant}
 
\renewcommand{\geq}{\geqslant}

\newcommand{\script}[1]{{\mathscr #1}}

\renewcommand{\Bbb}{\mathbb}
 
\newcommand{\N}{{\Bbb N}}
\newcommand{\R}{{\Bbb R}}

\newcommand{\Tref}[1]{Theo\-rem\,\ref{#1}}

\newcommand{\Cref}[1]{Co\-ro\-lla\-ry\,\ref{#1}}

\newcommand{\Ftwo}{{{\Bbb F}}_{\!2}}

\newcommand{\zero}{{\mathbf 0}}

\usepackage{stmaryrd} 
\usepackage{verbatim} 

\newcommand{\bigoh}{{\mathcal{O}}}
\usepackage{mathdots}
\usepackage{nicefrac}
\newcommand{\bit}{\ensuremath{\mathset{0,1}}}
\usepackage{float}
\usepackage[justification=centering]{caption} 
\usepackage{subcaption} 
\newcommand{\RNum}[1]{\uppercase\expandafter{\romannumeral #1\relax}}
\usepackage{enumerate}
\usepackage{enumitem} 

\usepackage{hyperref}
\hypersetup{
    colorlinks=true,
    linkcolor=black,
    urlcolor=black,
    linktoc=all,
    citecolor=black
}
\usepackage{adjustbox}

\IEEEoverridecommandlockouts

\begin{document}

\title{Capacity-achieving Polar-based Codes with Sparsity Constraints on the Generator Matrices}

\author{
James Chin-Jen Pang, Hessam Mahdavifar, and S. Sandeep Pradhan  
\thanks{This work was supported by the National Science Foundation under grants CCF--1763348, CCF--1909771, and 
CCF-2132815. 
This paper was presented in part at the 2020 IEEE International Symposium on Information Theory.
}%
\thanks{The authors are with the Department of Electrical Engineering and Computer Science, University of Michigan, Ann Arbor, MI 48109, USA (e-mail: cjpang@umich.edu; hessam@umich.edu; pradhanv@umich.edu).}
}

\maketitle
\vspace{-15mm}
 
\begin{abstract}
In general, the generator matrix sparsity is a critical factor in determining the encoding complexity of a linear code. 
Further, certain applications, e.g., distributed crowdsourcing schemes utilizing linear codes, require most or even all the columns of the generator matrix to have some degree of sparsity.
In this paper, we leverage polar codes and the well-established channel polarization to design capacity-achieving codes  with a certain constraint on the weights of all the columns in the generator matrix (GM) while having a low-complexity decoding algorithm.
We first show that given a binary-input memoryless symmetric (BMS) channel $W$ and a constant $s \in (0, 1]$, there exists a polarization kernel such that the corresponding polar code is capacity-achieving with the \textit{rate of polarization} $s/2$, and the GM column weights being bounded from above by $N^s$. 
To improve the sparsity versus error rate trade-off, we devise a column-splitting algorithm and two coding schemes for BEC and then for general BMS channels. The \textit{polar-based} codes generated by the two schemes inherit several fundamental properties of polar codes with the original $2 \times 2$ kernel including the  decay in error probability, decoding complexity, and the capacity-achieving property. Furthermore, they demonstrate the additional property that their GM column weights are bounded from above sublinearly in $N$, while the original polar codes have some column weights that are linear in $N$.
In particular, for any BEC and $\beta <0.5$, the existence of a sequence of capacity-achieving polar-based codes where all the GM column weights are bounded from above by $N^\lambda$ with $\lambda \approx 0.585$, and with the error probability bounded by $\bigoh(2^{-N^{\beta}} )$ under a decoder with complexity $\bigoh(N\log N)$, is shown. 
The existence of similar capacity-achieving polar-based codes with the same decoding complexity is shown for any BMS channel and $\beta <0.5$ with  $\lambda \approx 0.631$.
\end{abstract}

\section{Introduction}

Capacity-approaching error-correcting codes such as low-density parity-check (LDPC) codes \cite{gallager1962low} and polar codes \cite{arikan2009channel} have been extensively studied for applications in wireless and storage systems. Besides conventional applications of codes for error correction, a surge of new applications has also emerged in the past decade including crowdsourcing \cite{karger2011iterative,vempaty2014reliable}, distributed storage \cite{dimakis2010network}, and speeding up distributed machine learning \cite{lee2018speeding, SGDviaLDGM2019}. To this end, new motivations have arisen to study codes with sparsity constraints on their generator and/or parity-check matrices.
For instance, the stored data in a failed server needs to be recovered by downloading data from a few servers only, due to bandwidth constraints, imposing sparsity constraints in the decoding process in a distributed storage system. In crowdsourcing applications, e.g., when workers are asked to label items in a dataset, each worker can be assigned only a few items due to capability limitations, imposing sparsity constraints in the encoding process. 
More specifically, codes defined by sparse generator matrices become relevant for such applications \cite{mazumdar2017semisupervised,pang2019coding}. 

In this paper, we focus on polar codes in order to construct a sequence of codes defined by sparse GMs with practical utility, such as low decoding complexity, explicit construction, sufficiently fast decay in the error probability, and the potential to approach capacity at large block-length.

\subsection{Polar Codes}\label{subsec:Intro_polar}

Channel polarization, introduced by  Ar{\i}kan \cite{arikan2009channel, ArikanRatePolarization09}, is one of the most recent breakthroughs in coding theory. 
Polar codes are a class of provably capacity-achieving channel codes with explicit construction for general BMS channels, and have attracted significant attention due to their error correction performance, as well as their low-complexity decoding algorithms.
Within the ongoing fifth generation wireless systems (5G) standardization process, polar codes have been adopted for uplink and downlink control information for the enhanced mobile broadband (eMBB) communication service. 
Furthermore, polar codes and polarization phenomenon have been successfully applied to a wide range of problems including data compression~\cite{Arikan2,abbe2011polarization}, broadcast channels~\cite{mondelli2015achieving,goela2015polar}, multiple access channels~\cite{STY,MELK}, physical layer security~\cite{MV,andersson2010nested}, and coded modulations \cite{mahdavifar2015polar}. 

\subsection{LDGM and Related Works}

A related line of work on studying linear codes with sparsity constraints on their generator matrices is by associating them with sparse graph representations \cite{chandar2010sparse}. In this context, they are referred to as low-density generator matrix (LDGM) codes, also regarded as the counterpart of LDPC codes.
The sparsity of the generator matrices of LDGM codes leads to a low encoding complexity, and has been adopted in applications such as lossy source compression \cite{golmohammadi2018encoding} and multiple description coding \cite{zhang2012ldgm}.
In \cite{mackay1999good, mackay1995good} it was pointed out that certain constructions of LDGM codes are not asymptotically \textit{good}, a behavior which is also studied using an error floor analysis in \cite{zhong2005approaching, garcia2003approaching}. 

In terms of the sparsity of the GM, the authors of \cite{LDGM_capAchieving2011} showed the existence of capacity-achieving codes over binary symmetric channels (BSC) using random linear coding arguments when the column weights of the GM are upper bounded by $\epsilon N$, for any $\epsilon > 0$, where $N$ is the code block length.
Also, it is conjectured in \cite{LDGM_capAchieving2011} that column weight upper bounds that scale sublinearly in $N$ suffice to achieve the capacity. For binary erasure channels (BEC), bounds that scale as $\bigoh(\log N)$ suffice for achieving the capacity, again using random linear coding arguments \cite{LDGM_capAchieving2011}. 
Furthermore, the scaling exponent of such random linear codes are studied in \cite{mahdavifar2017scaling}.
Later, in \cite{lin2018coding}, the existence of capacity-achieving systematic LDGM ensembles over any BMS channel with the expected value of the weight of the entire GM bounded by $\epsilon N^2$, for any $\epsilon > 0$, is shown.
While the (ensemble-averaged) block-error probability for the codes goes to zero as the block-length grows large, the speed of decay in the error probability is not provided in \cite{LDGM_capAchieving2011, lin2018coding}.

In \cite{mazumdar2017semisupervised}, the problem of label learning through queries from  a crowd of workers was formulated as a coding theory problem. Due to practical constraints in such crowdsourcing scenarios, each query can only contain a small number of items. 
In \cite{pang2019coding}, we considered the same setting as in \cite{mazumdar2017semisupervised} with the additional consideration that some workers may not respond to queries, a scenario that resembles a binary erasure channel. Then we showed that a combination of LDPC codes and LDGM codes gives a query scheme where the number of queries approaches the information-theoretic lower bound \cite{pang2019coding}. 

In the realm of quantum error correction, quantum low-density-generator-matrix (QLDGM) codes, quantum low-density-parity-check (QLDPC) codes, and other sparse-graph-based schemes have been extensively studied due to  the small numbers of quantum interactions per qubit during the encoding and/or error correction procedure, avoiding additional quantum gate errors and facilitating fault-tolerant decoding.
Amongst these schemes, the error correction performance of the LDGM-based codes proposed in \cite{PhysRevA.102.012423} was shown to outperform all other Calderbank-Steane-Shor (CSS) and non-CSS codes of similar complexity. 

In both applications highlighted above, the benefit of the LDGM codes follows from a certain upper bound on the column weights of the GM, ensuring the columns are relatively low weight. 
Motivated by these applications, the main goal of this work is to construct sequences of codes where all of the columns of the GM are \textit{low weight}, where certain upper bounds on the weight will be specified later.

\subsection{Our Contributions}
In this paper, we study capacity-achieving polar and polar-based codes over BMS channels with sparsity constraints on generator matrix column weights. 
Leveraging polar codes based on general kernels, with rates of polarization studied in \cite{ korada2010polar}, we show that capacity-achieving polar codes with column weights bounded from above by $N^s$ exist for any given $s>0$, where $N$ is the code block length. This verifies the conjecture given in \cite{LDGM_capAchieving2011}.
There is, however, a trade-off between the sparsity parameter $s$ and the rate of polarization, given by $\frac{s}{2}$, and the decoding time complexity.

For the case when the speed of decay for block-error probability and the GM sparsity are both constrained, we propose two new code constructions with sparse GM columns, which provide a better trade-off for $s> 0.585$. 
We first consider BEC, and propose a splitting algorithm  termed \textit{decoder-respecting splitting (DRS)} algorithm, which, roughly speaking, splits \textit{heavy} columns in the GM into several \textit{light} columns.
Note that if one splits the heavy columns in an arbitrary manner to form a new GM, the code defined by the new GM may be substantially different from the original one in terms of the error probability and/or having a low-complexity decoder.
Leveraging the fact that the polarization transform of a BEC leads to BECs, the DRS algorithm converts the encoder of a standard polar code into an encoder defined by a sparse GM without hurting the reliability of the bit-channels observed by the source bits. Furthermore, the specific structure of DRS enables a low-complexity successive cancellation decoder in a recursive fashion inheriting that of original polar codes. 
In particular, we show a sequence of codes defined by GMs with column weights upper bounded by $N^{\lambda}$, for any $\lambda > \lambda^* \approx 0.585$ and the existence of a decoder with computation complexity $\bigoh(N \log N)$ under which the block-error probability is bounded by $2^{-N^{\beta}}$ for any $\beta < 0.5$.

Next, for general BMS channels, we propose an enhancement of the DRS-based encoding scheme, referred to as \textit{augmented-DRS (ADRS)} scheme, which requires additional channel uses and decoding complexity.  
In spite of these limitations, we show that there exists a sequence of capacity-achieving codes, referred to as the \textit{polar-ADRS} codes.
The sequence  of codes is defined by GMs with column weights upper bounded by $N^{\lambda}$, for any $\lambda > \lambda^{\dagger} \approx 0.631$, and can be decoded with complexity $\bigoh(N \log N)$.

The rest of the paper is organized as follows. In Section \ref{sec:Prelim}, we introduce basic notations and definitions for channel polarization and polar codes. 
Section \ref{sec:SparseLDGM} provides a sparsity result for polar codes with general kernels. 
In Section \ref{sec:LDGMwithDEC}, we introduce the DRS algorithm and the ADRS scheme, and the corresponding code constructions over the BEC and BMS channels, respectively. The successive cancellation decoders are also described and shown to be of low computation complexity.
Finally, Section \ref{sec:Conclusion} concludes the paper. 
The proofs for the results in Sections \ref{sec:SparseLDGM} and \ref{sec:LDGMwithDEC} are included in the Appendix.

\section{Preliminaries}\label{sec:Prelim}
Let $h_b(\cdot)$ denote the binary entropy function, $\exp_2{(x)}$ denotes the function $2^x$,  $\ln(\cdot)$ be the logarithmic function with base $e$,  
and $\log(\cdot)$ be the logarithmic function with base $2$. $Z(W)$ denote the Bhattacharyya parameter of a channel $W$. 
We give formal definitions for the BMS channel and capacity-achieving codes for readers' reference.
\begin{definition}\label{def:BMS}
A binary memoryless symmetric channel (BMS)  $W: \mathcal{X} \rightarrow \mathcal{Y} $ is a noisy memoryless channel with binary input alphabet $ \mathcal{X}$, and channel output alphabet $ \mathcal{Y}$, (we use $ \mathcal{X}  = \mathset{0, 1}$, and assume $ \mathcal{Y}$ is finite, in this paper.) such that $\Pr[Y = y\vert X=0] = \Pr[Y = \phi(y)\vert X=1]$ for all $y\in \mathcal{Y}$ for some involution $\phi$ on $\mathcal{Y}$.
\end{definition}

\begin{definition}\label{def:cap-achieving}
    A type of code is said to be capacity achieving over a BMS channel $W$ with capacity $C= I(W) > 0$ if, for any given constant $R <C$, there exists a sequence of codes  with rate $R$ and the block-error probability vanishes as the block length $N$ grows large. 
    The block-error probability is evaluated under the maximum likelihood (ML) decoder, unless a different decoding scheme is specified.
\end{definition}

\subsection{Channel Polarization and Polar Codes}\label{Prelim:polar}

The \emph{channel polarization} phenomenon was discovered by Ar{\i}kan \cite{arikan2009channel} and is based on the polarization transform as the building block.
Let $\mathcal{W}$ denote the class of all BMS channels.
The channel transform $W \mapsto (W^{-}, W^{+})$ that maps $\mathcal{W}$ to $\mathcal{W}^2$, where $W^{-} :\mathcal{X} \rightarrow  \mathcal{Y}^2$ and $W^{+} :\mathcal{X} \rightarrow  \mathcal{Y}^2 \times \mathcal{X}$, is defined in \cite{arikan2009channel} and is often referred to as a polarization recursion.
Then a channel $W^{s_1, s_2, \ldots , s_n}$ with $s_i \in \mathset{-,+}, i= 1,2,\ldots, n$, can be defined by applying the channel transform $n$ times recursively, as in \cite{arikan2009channel}.
 
For $N=2^n$, the polarization transform is obtained from the $N \times N$ matrix $G_2^{\otimes n}$, where $G_2={\tiny\begin{bmatrix}
		1 & 0 \\
		1 & 1 \\
		\end{bmatrix}}$~\cite{arikan2009channel}, and $A^{\otimes n}$ denote the $n$-fold Kronecker product of $A$.
A polar code of length $N$ is constructed by selecting certain rows of $G_2^{\otimes n}$ as its generator matrix.
More specifically, let $K$ denote the code dimension. Then all the $N$ bit-channels in the set $\{W^{s_1, s_2,\ldots, s_n}:$ $s_i\in \mathset{-,+} \mbox{ for } i =1,2,\ldots, n\}$, resulting from the polarization transform, are sorted with respect to an associated parameter, e.g., their probability of error (or Bhattacharyya parameter), the best $K$ of them with the lowest probability of error are selected, and then the corresponding rows from $G_2^{\otimes n}$ are selected to form the GM. 
Hence, the GM of an $(N,K)$ polar code is a $K \times N$ sub-matrix of $G_2^{\otimes n}$. Then the probability of error of this code, under successive cancellation (SC) decoding, is upper bounded by the sum of probabilities of error of the selected $K$ best bit-channels \cite{arikan2009channel}.

\subsection{General Kernels and Error Exponent}\label{Prelim:polarExp}
It is shown in \cite{korada2010polar} that if $G_2$ is replaced by an $l \times l$ matrix $G_l$, then polarization still occurs if and only if $G_l$ is an invertible matrix in $\mathbb{F}_2$ and it is not upper triangular under any column permutation, in which case the matrix $G_l$ is called a \textit{polarization kernel}. Furthermore, the authors of \cite{korada2010polar} provided a general formula for the error exponent of polar codes constructed based on an arbitrary $l\times l$ polarization kernel $G_l$. More specifically, let $N= l^n$ denote the block length and $C$ denote the capacity of the channel. For any fixed $\beta < E(G_l)$ and fixed code rate $R< C$, where  $E(G_l)$ denotes the rate of polarization (see \cite[Definition 7]{korada2010polar}), there is a sequence of polar codes based on $G_l$ with probability of error $P_e$ under SC decoding bounded by 
$
P_e(n) \leq 2^{-N^{\beta}},
$
for all sufficiently large $n$.  
The rate of polarization $E(G_l)$ is given by 
$ 	E(G_l) = \frac{1}{l}\sum_{i=1}^l \log_l D_i, $
where $\mathset{D_i}_{i=1}^l$ are the \textit{partial distances} of $G_l$. 
More specifically, for $G_l= [g_1^T, g_2^T, \ldots, g_l^T]^T$, the partial distances $D_i$ are defined 
given by $D_i \triangleq d_H(g_i, \mbox{span}(g_{i+1}, \ldots, g_l) )$ for $i = 1, 2, \ldots, l$,
where $d_H(a,b)$ is the Hamming distance between two vectors $a$ and $b$, and $d_H(a, U)$ is the minimum distance between a vector $a $ and a subspace $U$, i.e., $d_H(a, U)= \min_{u\in U}d_H(a,u)$.

\section{Sparse Polar Code Constructions based on Large Kernels}\label{sec:SparseLDGM}

In this section we first show the existence of capacity-achieving polar codes with generator matrices for which all column weights scale at most polynomially with arbitrarily small degree in the block length $N$, hence validating
the conjecture in \cite{LDGM_capAchieving2011}. 
Second, we show that, for any polar code of rate $1$, \textit{almost} all of the column weights of the GM are polynomial in $N$.
\begin{theorem}\label{prop:polarWithPolyColumnWeight}
	For  any fixed $s\in (0,1)$ and any BMS channel, there are capacity-achieving polar codes under SC decoding, with generator matrices having column weights bounded by $N^s$, where $N$ denotes the block length of the code.
\end{theorem} 

\proof
Consider an $l\times l$ polarizing matrix
$		{\small G_l=
		\begin{bmatrix}
		I_{\frac{l}{2}} & 0_{\frac{l}{2}} \\
		I_{\frac{l}{2}} & I_{\frac{l}{2}} \\
		\end{bmatrix}},
		$
	where $l$ is an even integer such that $l \geq 2^{\frac{1}{s}}$.
The partial distances are
$ D_i= 1$ for $1\leq i\leq \frac{l}{2}$ and $D_i= 2$ for $\frac{l}{2}+1 \leq i\leq l$. Hence, the rate of polarization 
	$E(G_l) = \frac{1}{2}\log_l 2  >0 $, and there is a sequence of capacity-achieving polar codes constructed using $G_l$ as the polarizing kernel. Note that in $G_l$, each column has weight at most $2$ and, hence, the column weights of $G_l^{\otimes n}$ are upper bounded  by $2^n$. By the specific choice of $l$, we have
    $
    2^n \leq {(l^s)}^n = {(l^n)}^s = N^s, $
    where $N= l^s$ is the block length of the code. This completes the proof. 
\endproof

\begin{remark}\label{rem:poly_sparse_polar}
While \Tref{prop:polarWithPolyColumnWeight} provides a theoretical guarantee on the existence of capacity-achieving polar codes with sparse generator matrices, the sparsity comes at a cost. 
Specifically, the rate of polarization $E(G_l) = \frac{1}{2}\log_l 2  \leq \frac{s}{2}$ is smaller than that associated with the kernel $G_2$, given by $E(G_2) = 0.5$.
On the other hand, while the SC decoding complexity for polar codes defined by general $l\times l$ kernels behaves as $\bigoh(\frac{2^l}{l} N \log_l{N})$\cite{korada2010polar}, in this case, the complexity scales as $O(N \log_l N)$ by considering the following viewpoint.  Interleave $(l/2)^n$ copies of the polar code with block length $2^n$ based the standard $G_2$ kernel, to form a code with block length $N = l^n$ with an $n$-stage recursive encoder structure.  By decoding each copy with complexity $O(2^n \log 2^n) = O(n 2^n)$ under the SC decoder, the entire code can be decoded with complexity $O((l/2)^n \cdot n 2^n) = O(N \log_l N)$.
\end{remark}

Since we can construct capacity-achieving codes with column weights upper bounded by $N^s$ with any fixed $s>0$, by using polar codes, the question now is whether it is possible to further improve the sparsity of polar code GMs. For instance, we know it is possible to have an upper bound of $\bigoh(\log N)$ on all the GM column weights of capacity-achieving codes, over the BEC, by utilizing random linear ensembles \cite{LDGM_capAchieving2011}. 
For rate-$1$ polar codes, the proposition below answers the inquiry in the negative, by showing that almost all the GM columns have weights lower bounded by a polynomial in $N$.
\begin{proposition}\label{prop:noLogColumns}
	Given any $l \geq 2$, $l\times l $  polarizing kernel $G_l$, and $\frac{1}{l} > r > 0$, 
    the fraction of columns in $G_l^{\otimes n}$ with $\bigoh(N^{r \log_l{2} } )$ Hamming weight vanishes as $n$ grows large, where $N = l^n$.
\end{proposition}
\proof
The proof is given in Appendix Section\ref{pf:sec:SparseLDGM}. \endproof

The trade-off highlighted in Remark \ref{rem:poly_sparse_polar} suggests that off-the-shelf polar code constructions with large kernels may not be the ideal option when the speed of decay of the error probability is a concern. However, the heaviest column in the polar code with kernel $G_2$ scales as $\Theta(N)$ for any code rate. 
To construct codes with sparse GM and suitable decay of the error probability, in the next section, we propose a \textit{splitting} algorithm for the generator matrix and investigate the resulting codes in terms of the error probability, GM column sparsity, and the decoding complexity.

\section{Sparse Polar-based Codes with Low-complexity Decoding}\label{sec:LDGMwithDEC}
When all columns of a matrix $G$ are required to be sparse, that is, have low Hamming weights, a \textit{splitting algorithm} is applied. 
Given a column weight threshold  $w_{u.b.}$, a splitting algorithm splits any column in $G$ with weight exceeding $w_{u.b.}$ into columns that  sum to the original column both in $\Ftwo$ and in $\R$, and that have weights no larger than $w_{u.b.}$. 
Note that a column of $G$ is left intact by the splitting algorithm as long as its Hamming weight does not only exceed $w_{u.b.}$. Thus a splitting algorithm would be described as an algorithm which takes as input a column vector $\mathbf{v}$ and a weight threshold  $w_{u.b.}$, and returns a set of column vectors whose lengths are equal to the length of $\mathbf{v}$. 
Given a matrix $A$ with $m$ columns and a threshold  $w_{u.b.}$, with a slight abuse of notation, the matrix \textit{generated} by a splitting algorithm is defined as the matrix whose column vectors are those from the $m$ sets, which are respectively the outputs of the algorithm for each column of $A$. For example, consider a $4\times 2$ matrix $A = 
   \small{ \begin{bmatrix}
		1 & 0 & 1 & 1 \\
		1 & 1 & 1 &0 \\
    \end{bmatrix}^T } = [\mathbf{a_1}, \mathbf{a_2}]
$, a threshold $w_{u.b.} = 2$, and a splitting algorithm $\cS$. 
Let $\cS(\mathbf{a_i}, w_{u.b.})$, $i= 1,2$, be the sets of vectors returned by  $\cS$, given by 
$    \cS(\mathbf{a_1}, w_{u.b.}) = \mathset{[1, 0, 1, 0]^T,[0, 0, 0, 1]^T }, 
    \cS(\mathbf{a_2}, w_{u.b.}) = \mathset{[1, 0, 1, 0]^T,[0, 1, 0, 0]^T }.$ 
The matrix {generated} by  $\cS$ for $A$ is then a $4\times 4$ matrix of the form $[
[1, 0, 1, 0]^T,[0, 0, 0, 1]^T, [1, 0, 1, 0]^T,[0, 1, 0, 0]^T 
]$, or a column permutation of it.

Let an $(N,K)$ polar code $\cC$ have a $K\times N$ submatrix of $G = G_2^{\otimes n}$ as the generator matrix, and 
$G'$ denote the $N\times N(1+\gamma)$ matrix generated by the splitting algorithm, where $N= 2^n$.  
A new code\textit{ based on} $G'$ selects the same $K$ rows of $G'$ as the polar code $\cC$ to form the generator matrix, where all the column weights are bounded by $w_{u.b.}$. 
Such a code is referred to as a \textit{polar-based code corresponding to} $G'$, or a PB$(G')$ code, in this paper.

Note that more detailed description is needed to uniquely specify a splitting algorithm, which then determines the term $\gamma$ and the performance of the PB$(G')$ code.
Specifically, the channel polarization phenomenon and the recursive encoding and decoding structure may not be valid when the GM is modified by the splitting algorithm. 
These changes also imply that the codes with the split GM may suffer from drawbacks such as weaker bounds on error probability and larger decoding complexity, as well as the rate loss with a multiplicative factor of $1+ \gamma$, when compared to the polar codes. 
In this section, we introduce a \textit{splitting} algorithms, referred to as the decoder-respecting splitting (DRS) algorithm, and two encoding schemes that are effective in avoiding the drawbacks. These schemes enable low-complexity SC decoders based on likelihood ratios that can be calculated with a recursive algorithm. 
Specifically, when the threshold  $w_{u.b.}$ is chosen appropriately, we show in Section~\ref{sub:DRSalg} that the term $\gamma$ goes to $0$ exponentially fast in $n$, when the DRS algorithm is applied to columns of the matrix $G= G_2^{\otimes n}$. 
The encoding of the resulting PB$(G')$ codes can be realized by a encoding scheme which inherits the recursive structure of the original polar codes, except only at locations that corresponds to a split of a column of $G$, as dictated by the DRS algorithm. At these locations, the exclusive-OR operations are removed and additional copies of the underlying channel are used.
The PB$(G')$ codes suffer only a negligible $1+\gamma$ multiplicative factor of rate loss compared to the original polar codes for large $n$. For BEC, this sequence of codes is capacity-achieving with an error exponent of $\frac{1}{2}$, under a new SC decoding scheme (see \Tref{Thm:BEClowcomplDEC} in Section~\ref{subsection:LCdec_BEC}).
For general BMS channels, another encoding scheme, referred to as the \textit{ADRS} scheme, is proposed in Section~\ref{subsection:LCdec_BMS}. This scheme introduces  additional `noise' nodes and requires even more copies of the underlying channel when the DRS algorithm requires a split. For codes generated by this scheme, results similar to that in \Tref{Thm:BEClowcomplDEC} are available with a slightly stricter condition on the choice of $w_{u.b.}$.

\subsection{Decoder-Respecting Splitting Algorithm} \label{sub:DRSalg}

The main idea of the DRS algorithm is to construct a generator matrix that can be realized with an encoding pattern similar to conventional polar codes such that the column weights of the matrix associated with the diagram are at most $w_{u.b.}$. 
The pseudo code for the algorithm is provided in Algorithm~\ref{alg1}.

\begin{algorithm} 
\small
\caption{DRS algorithm} 
\label{alg1} 
\textbf{Input:} weight threshold $w_{u.b.}\in \N$, a column vector $\mathbf{v}\in \bit^{2^n \times 1}$    \\
\textbf{Output:} the set of vectors with length $2^n$ returned by \textproc{DRS-Split}($w_{u.b.}, \mathbf{v}$)

\begin{algorithmic}[1]
\Function{DRS-Split}{$w_{u.b.}, \mathbf{x}$}
    
    \If{$w_H(\mathbf{x})> w_{u.b.}$}
        \State $k\gets \mbox{length}(\mathbf{x}) /2$
        \State $\mathbf{x}_h \gets (x_1,\ldots, x_k)^T$, $\mathbf{x}_t \gets (x_{k+1},\ldots, x_{2k})^T$
        \State $Y_h \gets$ \textproc{DRS-Split}($w_{u.b.}, \mathbf{x}_h$)
        \State $Y_t \gets$ \textproc{DRS-Split}($w_{u.b.}, \mathbf{x}_t$)
        \If{$\mathbf{x}_h= \zero_{k\times 1} $} 
            \State \Return $\bigcup\limits_{y\in Y_t} \{(\zero_{1\times k}, y^T)^T \}$ 
        \ElsIf{$\mathbf{x}_t= \zero_{k\times 1} $} 
            \State \Return $\bigcup\limits_{y\in Y_h} \{ (y^T, \zero_{1\times k})^T \}$ 
        \Else 
            \State \Return $\bigcup\limits_{y\in Y_t} \{ (\zero_{1\times k}, y^T)^T\} \cup  \bigcup\limits_{y\in Y_h} \{(y^T, \zero_{1\times k})^T \}$ 
        \EndIf
    \ElsIf{$w_H(\mathbf{x})=0 $}
        \State \Return \{\}
    \Else
        \State \Return $\mathset{ \mathbf{x}}$
    \EndIf
\EndFunction
\end{algorithmic}
\end{algorithm}
The core of the algorithm is the \textproc{DRS-Split} function. 
When the weight of the input the vector $\mathbf{x}$ is larger than the threshold, it splits the vector in half into vectors $\mathbf{x}_h$ and $\mathbf{x}_t$, and recursively finds two sets, $Y_h$ and $Y_t$, composed of vectors with the length halved compared to the length of $\mathbf{x}$. 
The vectors are then appended to the length of $\mathbf{x}$, which collectively form the output of the function.
For a vector $\mathbf{u}\in \bit^{m\times 1}$, let $\abs{\mathbf{u}} = m$ denote its length, and $w_H(\mathbf{u})$ its Hamming weight.
We note that the weights of vectors in  $Y_h$ and $Y_t$ are respectively upper bounded by the weights of $\mathbf{x}_h$ and $\mathbf{x}_t$, both of which are bounded by $k= \abs{ \mathbf{x}_h}=\abs{ \mathbf{x}_t}$, and that the value of $k$ is halved each iteration.
Hence, the function is guaranteed to terminate as long as the threshold is a positive integer.

We use a simple example to illustrate the algorithm. Let $n=3 $, $v= [0,0,0,0, 1,1,1, 1]^T$ and $w_{u.b.}= 2$. Since the weight of $v$ exceeds the threshold, it is first split into 
$\mathbf{x}_h=[0,0,0,0]^T$ and $\mathbf{x}_t= [ 1,1,1, 1]^T$.
Since $\mathbf{x}_h$ is an all-zero vector, $Y_h$ is an empty set according to line $14$ to $15$. 
To compute $Y_t=$\textproc{DRS-Split}($2, [ 1,1,1, 1]^T$), the function splits the input into half again, thereby obtaining $\mathbf{x}_h'=[1,1]^T$ and $\mathbf{x}_t'= [ 1,1]^T$.
The corresponding $Y_h'$ and $Y_t'$ are then both $\mathset{[1,1]^T}$ and, hence, we have $Y_t= \mathset{ [0,0, 1,1]^T}\cup \mathset{ [1,1,0,0]^T}= \mathset{ [0,0, 1,1]^T, [1,1,0,0]^T}$.  
Since $\mathbf{x}_h= \zero_{4\times 1}$, the function proceeds to lines $7$ and $8$, and returns $\mathset{ [0,0,0,0,0,0, 1,1]^T, [0,0,0,0,1,1,0,0]^T}$.

In order to analyze the effect of the DRS algorithm on the matrix $G_2^{\otimes n}$, 
we show that the size of the algorithm output does not depend on the order of a sequence of Kronecker product operations,  where the \textit{size} of a set of vectors stands for the number of vectors in the set.
Suppose that the Kronecker product operations with the vector $[1 ,\; 1]^T $ for $n_1$ times and  with the vector $[0 ,\; 1]^T $ for $n_2$ times are applied on a vector $v$, where $n =n_1+n_2$ and the order of the operations is specified by a sequence $(s_1, s_2, \ldots, s_{n} ) \in \mathset{-,+}^{n}$ with $\abs{ \mathset{ i:s_i = - } } = n_1$ and $\abs{ \mathset{ i:s_i = + } } = n_2$. 
Also, let $v^{(i)}$ denote the output of applying the first $i$ Kronecker product operations on $v$. It is defined by the following recursive relation: 
\begin{equation} \label{eq:vi_recursion}
    v^{(i)} = \begin{cases}
    v^{(i-1)} \otimes [1, 1]^T,\; \mbox{if } s_i= -, \\
    v^{(i-1)} \otimes [0, 1]^T,\; \mbox{if } s_i= + ,
    \end{cases}
\end{equation} for $i \geq 1 $ and the initial condition $v^{(0)}= v$. We use  $v^{(s_1, s_2, \ldots, s_i)}$ instead of $v^{(i)}$ when the sequence is needed for clarity.
The following lemma shows that any two vectors of the form $v^{(s_1, s_2, \ldots s_{n})}$ will be split into the same number of columns under the DRS algorithm as long as  the sequences associated with them contain the same number of $-$ and $+$ signs.

\begin{lemma}\label{lem:rateLossIndepOfOrder}
    Let $n= n_1+n_2$ and  $(s_1, s_2, \ldots, s_{n} ) \in \mathset{-,+}^{n}$ be a sequence with $n_1$ minus signs and $n_2 $ plus signs.
    Let $v^{(n)}$ be the vector defined by a vector $v$ and the sequence $(s_1, s_2, \ldots, s_{n} )$  through equation \eqref{eq:vi_recursion}. 
    Then the size of the DRS algorithm output for $v^{(n)}$ depends only on the values $n_1$ and $n_2$.  
\end{lemma}
\proof
The proof is given in Appendix Section \ref{pf:lem:rateLossIndepOfOrder}.
\endproof
    
Let a $K\times N$ matrix $M= [\mathbf{u}_1,\mathbf{u}_2,\ldots, \mathbf{u}_N] $ and a threshold $w_{u.b.}$ be given. 
Suppose that the DRS algorithm is applied to each column in $M$ and the sum of the sizes of the output sets is $N(1+\gamma)$. 
Then \textproc{DRS}$(M)$ is defined as the $K \times N(1+\gamma)$ matrix consisting of all the vectors in the output sets (with repetition). 
    
We study the effect of the DRS algorithm in terms of the multiplicative rate loss, i.e., $1+\gamma$. Since all the columns of $G_2^{\otimes n}$ are in the form of $v^{(s_1, s_2, \ldots, s_n)}$ with $v = [0, 1]^T$ or $v = [1, 1]^T$, Lemma~\ref{lem:rateLossIndepOfOrder} substantially simplifies the analysis for $\gamma$. In particular, the following proposition shows an appropriate choice of $w_{u.b.}$ guarantees the existence of a sparse polar-based GM with vanishing $\gamma$.
    
\begin{proposition}\label{prop:DRSrateloss}
    Let the columns of  $G_2^{\otimes n}$ be the inputs for the DRS algorithm and \textproc{DRS}$(G_2^{\otimes n})$ be the $N \times N(1+\gamma)$ matrix generated by the DRS algorithm for $G_2^{\otimes n}$. 
    The term $\gamma$ vanishes exponentially fast as $n$ goes to infinity for any $w_{u.b.} = 2^{n\lambda}$ with $\lambda > \lambda^* \triangleq h_b(\frac{2}{3}) -\frac{1}{3} \approx 0.585$.
\end{proposition}    
\proof
The proof is given in Appendix Section \ref{pf:prop:DRSrateloss}.
\endproof

\begin{figure}[h!]
     \centering
     \begin{subfigure}[b]{0.47\textwidth}
         \centering
         \includegraphics[width=\textwidth]{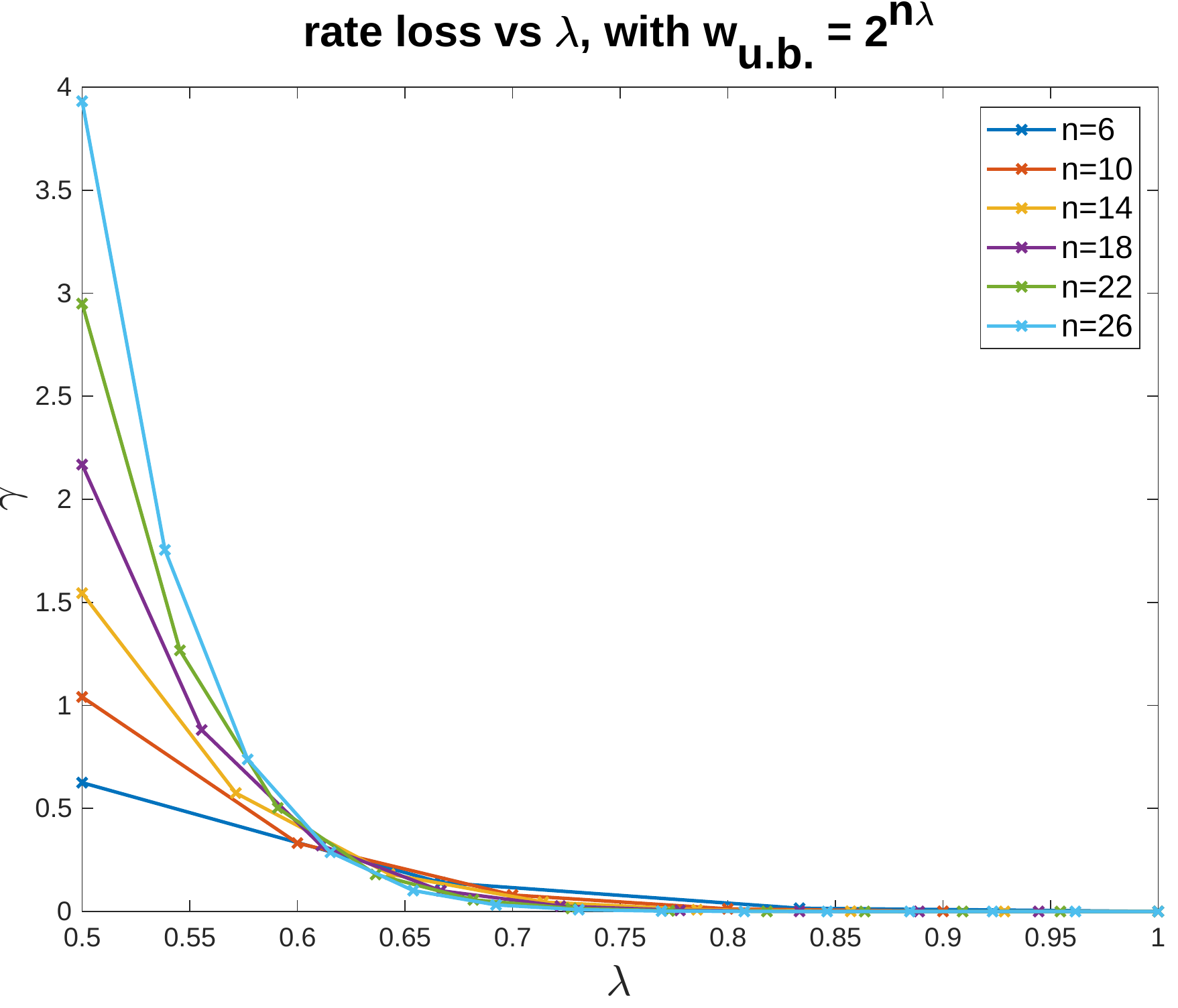}
         \caption{$\lambda \geq 0.5$}
        \label{fig:rateloss_sub1}
     \end{subfigure}
     \hfill
     \begin{subfigure}[b]{0.47\textwidth}
         \centering
         \includegraphics[clip, width=\textwidth]{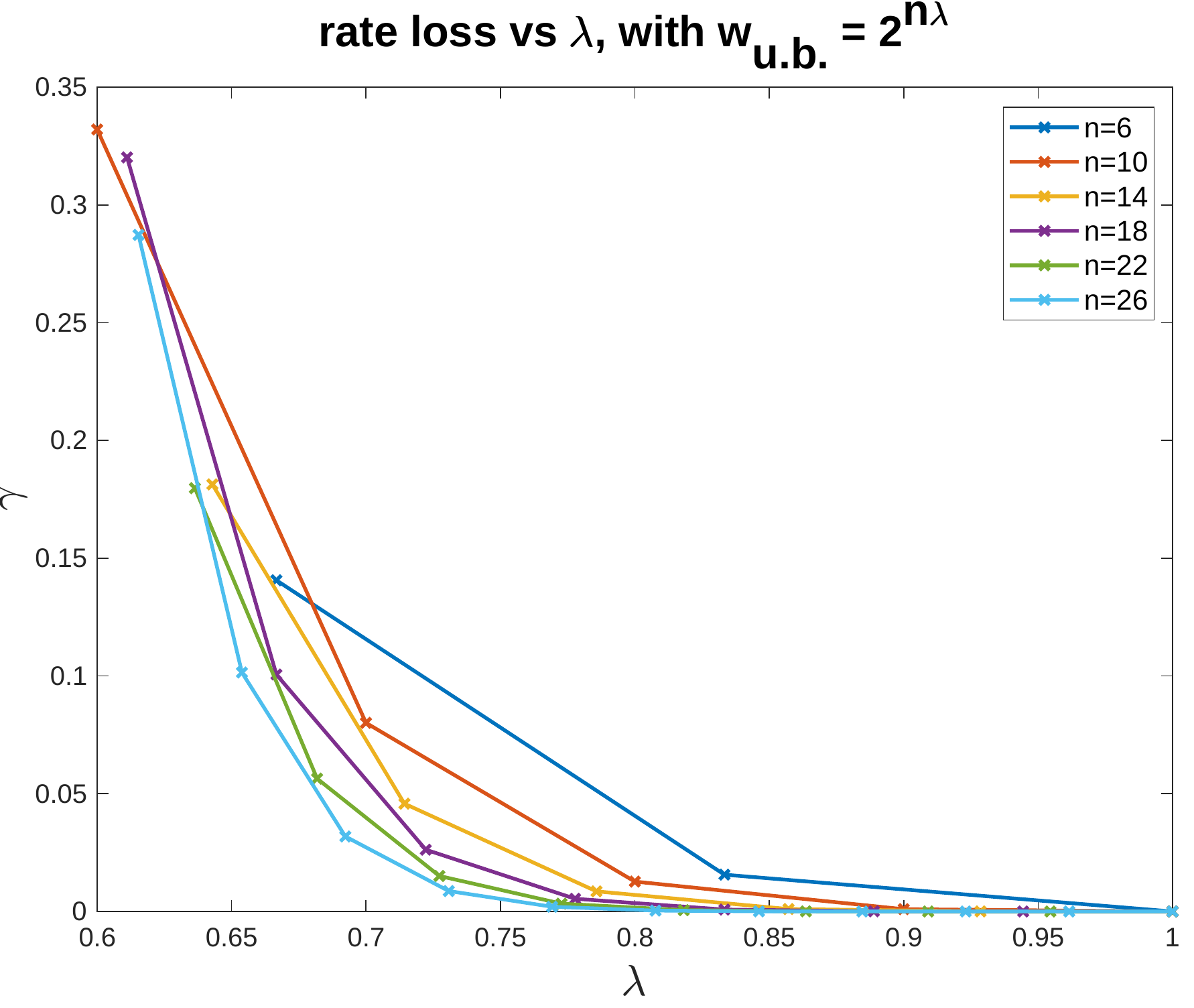}
        \caption{$\lambda \geq 0.585$}
    \label{fig:rateloss_sub2}
     \end{subfigure}
\caption{Multiplicative rate loss factor $\gamma$ versus $\lambda$, where $w_{u.b.} = N^{\lambda}$}
        \label{fig:rateloss_v_lambda_finite_n}
\vspace{-8mm}
\end{figure}

For the effect of the DRS algorithm on $G_2^{\otimes n}$ with finite $n$, we compute values of $\gamma$ for various combinations of $n$ and $\lambda$, as shown in Figures~\ref{fig:rateloss_sub1} and \ref{fig:rateloss_sub2}. The numerical results with $6\leq n \leq 26$ indicate that, for  $0.5 \leq \lambda < 0.6$, the multiplicative rate loss $\gamma$ is larger with larger $n$, and for $\lambda \geq 0.65$, $\gamma$ is smaller with larger $n$. The fact that the $n =26$ does not provide the smallest $\gamma$ for $\lambda$ close to $\lambda^*$ should not be considered a contradiction to Proposition~\ref{prop:DRSrateloss}. Instead, the closer $\lambda > \lambda^*$ is, the larger $n$ it takes for the exponential decay of $\gamma$ to dominate.

\subsection{Low-complexity Decoder for Polar-based Codes: BEC}
\label{subsection:LCdec_BEC}

In this section, we show two results for the polar-based code corresponding to  \textproc{DRS}($G_2^{\otimes n}$) over the BEC. 
Such codes are referred to as the \textit{polar-DRS} codes in this paper. 
First, we propose a low-complexity suboptimal decoder for the polar-DRS codes.
Second, with the low-complexity suboptimal decoder, the polar-DRS codes are capacity-achieving for suitable column weight threshold. 

It is known that when the channel transformation with kernel $G_2$ is applied to two BECs, the two new bit-channels are also BECs. 
Specifically, for two binary erasure channels $W_1 $ and $W_2$ with erasure probabilities $\epsilon_1 $ and  $\epsilon_2$, respectively, the polarized bit-channels $W^-(W_1, W_2) $ and $W^+(W_1, W_2) $ are binary erasure channels with erasure probabilities $\epsilon_1 +\epsilon_2 - \epsilon_1  \epsilon_2  $ and $\epsilon_1  \epsilon_2  $, respectively.

The mutual information $I(\cdot)$ and Bhattacharyya parameter $Z(\cdot)$ of a BEC $W$ with erasure probability $\epsilon $ are given by: 
        $I(W)= 1-\epsilon, 
        Z(W)= \epsilon.$
For a sequence $(s_1, s_2, \ldots, s_n) \in \mathset{-,+}^n$, the function  $\mathsf{Bi2De}(s_1, s_2, \ldots, s_n)$ returns the decimal value of the binary string in which a minus sign for $s_i$ is regarded as a $0$ and a plus sign as a $1$, e.g.,  $\mathsf{Bi2De}(-, +, +)= (011)_2 =3$.
Let $G$ denote $G_2^{\otimes n}$ and $G'$ denote  \textproc{DRS}($G_2^{\otimes n}$), and let $Z_{G}^{(s_1s_2\ldots s_n)}$ denote the Bhattacharyya parameter of the bit-channel $W^{s_1s_2\ldots s_n}$, which is equal to  $W_N^{(\mathsf{Bi2De}(s_1, s_2, \ldots, s_n)+1)}$  in \cite[page 3]{arikan2009channel}.
The term $Z_{G'}^{(s_1s_2\ldots s_n)}$ denotes the Bhattacharyya parameter of the bit-channel observed by the source bit of the same index corresponding to $G'$. 

The following lemma shows that the bit-channel observed by each source bit is better in terms of the Bhattacharyya parameter when $G'$ is the generator matrix instead of $G$.

\begin{figure}[h!]
     \centering
     \begin{subfigure}[b]{0.47\textwidth}
         \centering
         \includegraphics[trim=1.5cm 6.8cm 5.6cm 11.5cm, clip, width=\textwidth]{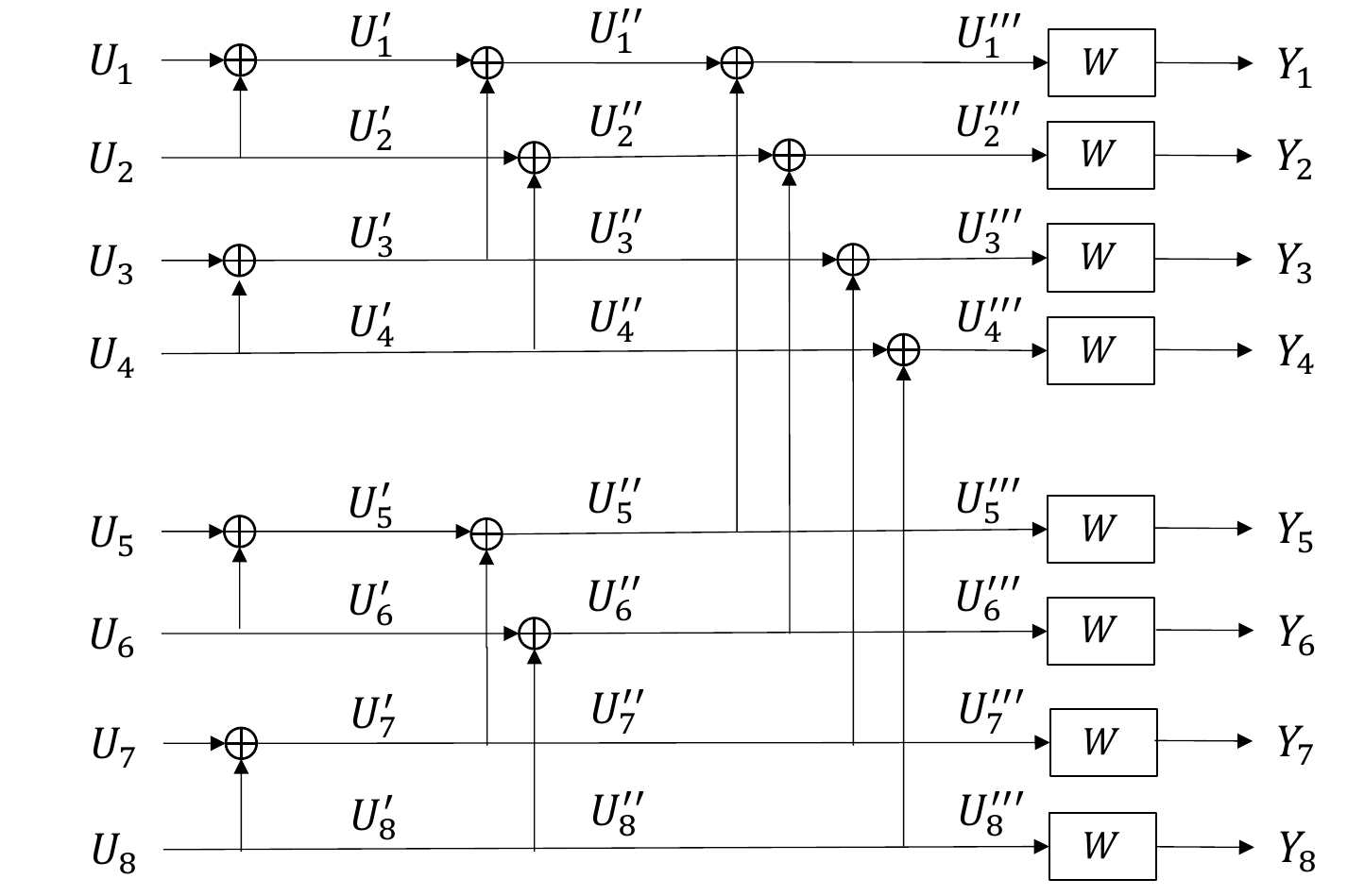}
         \caption{Encoding block for generator matrix $G= G_2^{\otimes 3}$}
        \label{fig:G2_BEC_EncoderN8}
     \end{subfigure}
     \hfill
     \begin{subfigure}[b]{0.5\textwidth}
         \centering
         \includegraphics[trim=2.4cm 6.9cm 3.8cm 11.5cm, clip, width=\textwidth]{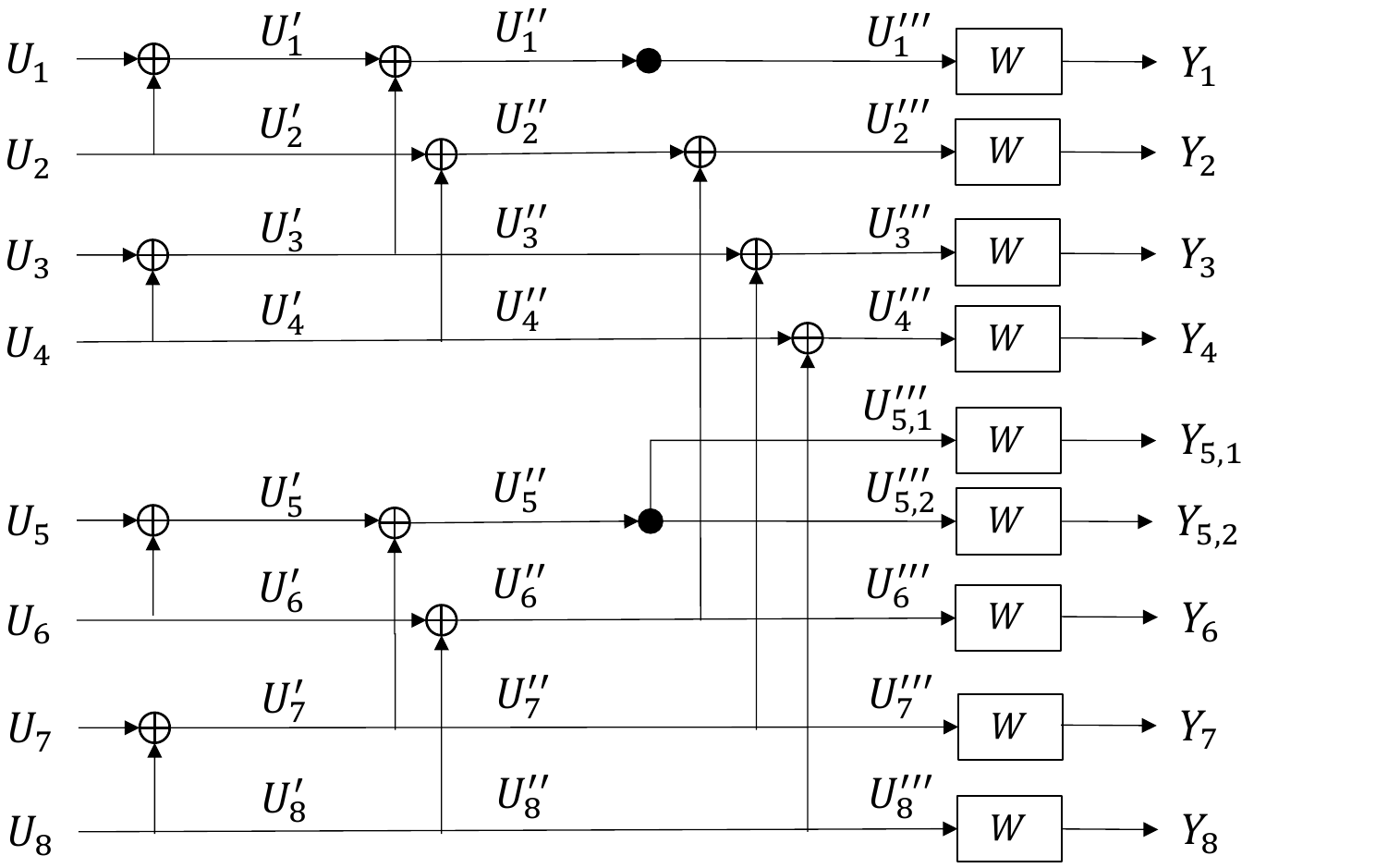}
        \caption{Encoding block for generator matrix $G'$}
    \label{fig:G2_BEC_EncoderN8w4}
     \end{subfigure}
\caption{}
        \label{fig:N8_BEC_1}
\vspace{-5mm}
\end{figure}

\begin{figure}[h!]
     \centering
     \includegraphics[trim=0.1cm 3.2cm 4.1cm 11.9cm, clip, width=\textwidth]{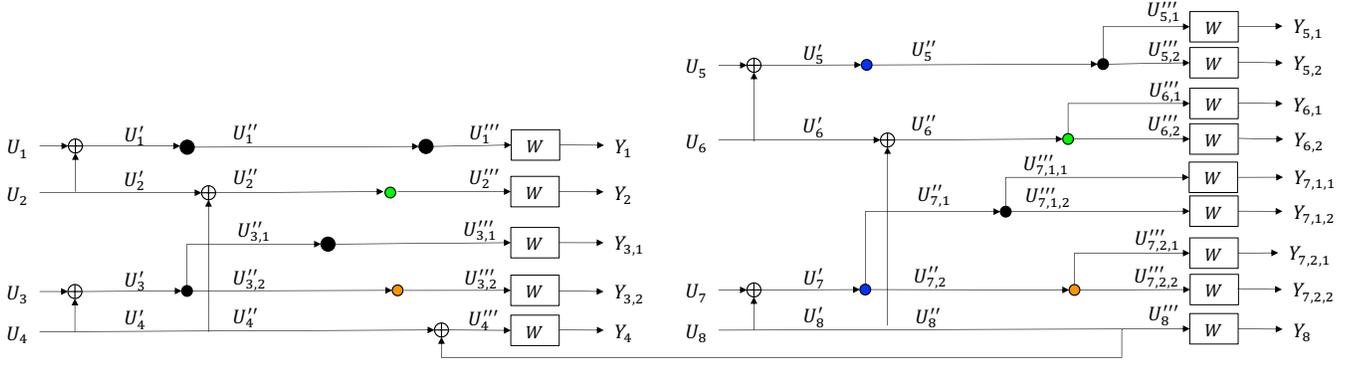}
     \caption{Encoding block for generator matrix \textproc{DRS}($G_2^{\otimes 3}$) when $w_{u.b.} =2$}
    \label{fig:N8_BEC_w2}
\vspace{-8mm}
\end{figure}

\begin{lemma}\label{lem:DRS_Z_BEC}
Let $w_{u.b.}$ and $n$ be given, and let $G$ denote $G_2^{\otimes n}$ and $G'$ denote  \textproc{DRS}($G_2^{\otimes n}$).
The following is true for any $(s_1, s_2, \ldots, s_n)\in \mathset{-,+}^n$:
    \[
    Z_{G'}^{(s_1s_2\ldots s_n)} \leq Z_{G}^{(s_1s_2\ldots s_n)}.
    \]
\end{lemma}
\proof
The proof is given in Appendix \ref{appendix:proof_LCdec_BEC}. 
\endproof
\begin{remark}\label{rem:BECencoder}

    A key to the proof of Lemma~\ref{lem:DRS_Z_BEC} is a recursive encoding scheme for the relationship $\mathbf{x} = \mathbf{u}G'$, where $ \mathbf{u}$ and $ \mathbf{x}$ are column vectors of lengths $N = 2^n$ and $N(1+\gamma)$, respectively. 
    The encoding scheme is most easily understood by considering the low-complexity encoding structure for the standard polar code, as seen in \cite{arikan2015origin}, and replacing the exclusive-OR (XOR) operations at locations that correspond to the splitting operations dictated by the DRS algorithm. 
    Specifically, when a split is required on a column of $G$, the corresponding XOR node is removed, the input bit for the `worse' channel remains untouched, and two copies of the input bit for the `good' channel are transmited through the underlying channel. 
    For example, consider $G = G_2^{\otimes 3}$. The encoding diagram for codes defined by $G$ is shown in Figure~\ref{fig:G2_BEC_EncoderN8}. We have 
    \[\def\arraystretch{0.6}{
    G= 
    \small\arraycolsep=0.5\arraycolsep\ensuremath{
    \begin{bmatrix}
1 & 0 & 0 & 0 & 0 & 0 & 0 & 0 \\
1 & 1 & 0 & 0 & 0 & 0 & 0 & 0 \\
1 & 0 & 1 & 0 & 0 & 0 & 0 & 0 \\
1 & 1 & 1 & 1 & 0 & 0 & 0 & 0 \\
1 & 0 & 0 & 0 & 1 & 0 & 0 & 0 \\
1 & 1 & 0 & 0 & 1 & 1 & 0 & 0 \\
1 & 0 & 1 & 0 & 1 & 0 & 1 & 0 \\
1 & 1 & 1 & 1 & 1 & 1 & 1 & 1 
    \end{bmatrix}}, 
    G' = \small\arraycolsep=0.9\arraycolsep\ensuremath{
    \begin{bmatrix}
1 & 0 & 0 & 0 & 0 & 0 & 0 & 0 & 0 \\
1 & 0 & 1 & 0 & 0 & 0 & 0 & 0 & 0 \\
1 & 0 & 0 & 1 & 0 & 0 & 0 & 0 & 0 \\
1 & 0 & 1 & 1 & 1 & 0 & 0 & 0 & 0 \\
0 & 1 & 0 & 0 & 0 & 1 & 0 & 0 & 0 \\
0 & 1 & 1 & 0 & 0 & 1 & 1 & 0 & 0 \\
0 & 1 & 0 & 1 & 0 & 1 & 0 & 1 & 0 \\
0 & 1 & 1 & 1 & 1 & 1 & 1 & 1 & 1 
    \end{bmatrix}}}, 
    \] where $G'$ is the matrix  \textproc{DRS}($G_2^{\otimes 3}$) when $w_{u.b.} =4$.
    The encoding structure for $G'$ is shown in Figure~\ref{fig:G2_BEC_EncoderN8w4}. 
    Since the first column is the only column of $G$ split by the DRS algorithm when $w_{u.b.} =4$, we remove the XOR node that performs $U_1''' = U_1'' + U_5''$, and assigns $U_1''' = U_1''$, $U_{5,1}''' = U_5''$ and $U_{5,2}''' = U_5''$. Two  solid circles, representing transparent nodes where the output variable(s) are identical to the input variable, are used to indicate the location of the removed XOR node. 
    For the case when $w_{u.b.} =2$, the DRS algorithm would split the first  column of $G$ into three vectors, and the second, third, and fifth column once each. 
    The corresponding encoding diagram is shown in Figure~\ref{fig:N8_BEC_w2}, where we color the solid circles associated with splits on the first, second, third, and fifth columns by black, green, orange, and blue, respectively.    

\end{remark}

We are ready to show the existence of a sequence of capacity-achieving codes over the BEC with GMs where the column weights are bounded by a polynomial in the blocklength, and that the block error probability under a low  complexity decoder vanishes as the $n$ grows large.
\begin{theorem}
     \label{Thm:BEClowcomplDEC}
    Let $\beta < E(G_2)=0.5 $, $\lambda >\lambda^*=h_b(\frac{2}{3}) -\frac{1}{3}\approx 0.585$, and a BEC $W$ with capacity $C$ be given. There exists a sequence of polar-based codes corresponding to  \textproc{DRS}($G_2^{\otimes n}$)
    with the following properties for all sufficiently large $n$:
       (1) The error probability under a SC decoder is upper bounded by $2^{-N^{\beta}}$, where $N= 2^n$,
        (2) The Hamming weight of each column of the GM is upper bounded by $N^\lambda$,
        (3) The rate approaches $C$ as $n$ grows large, and 
        (4) The codes can be decoded by a SC decoding scheme with complexity $O(N\log N)$.
\end{theorem}

\proof
Let the threshold for DRS algorithm be $w_{u.b.} = 2^{n\lambda}$,  $G$ denote $G_2^{\otimes n}$, and $G'$ denote \textproc{DRS}($G_2^{\otimes n}$).
We prove the four claims in order. 
First, Lemma \ref{lem:DRS_Z_BEC} shows that for a given $n$ and any  $t>0$, the following is true:
\be{eq:bhataPara_G'}
	\mathset{\mathbf{s}\in \mathset{-,+}^n :Z_{G}^{\mathbf{s}} \leq t  }
	\subseteq
	\mathset{
		\mathbf{s}\in \mathset{-,+}^n :Z_{G'}^{\mathbf{s}} \leq t  
	}.
\ee
Using \cite[Theorem 2]{arikan2009channel}, for any $\beta< \frac{1}{2}$, we have
\be{eq:rate_to_C}
\liminf_{n \to \infty} \frac{1}{N}
\abs{\mathset{\mathbf{s}\in \mathset{-,+}^n :Z_{G}^{\mathbf{s}} \leq 2^{-N^\beta}  }} = I(W) =C
\ee

Let $\mathbfsl{S_G} $ and $\mathbfsl{S_{G'}} $ denote the sets of the sequences $\mathbf{s} \in \mathset{-,+}^n$ that satisfy $Z_{G}^{\mathbf{s}} \leq 2^{-N^\beta}$ and $Z_{G'}^{\mathbf{s}} \leq 2^{-N^\beta}$, respectively.  Equation \eqref{eq:bhataPara_G'} guarantees that $\mathbfsl{S_G} $ is a subset of $\mathbfsl{S_{G'}}$.
Assume the code corresponding to $G$ freezes the input bits observing bit-channels $W^{s_1s_2\ldots s_n}$ for all $(s_1, s_2, \ldots, s_n) \notin \mathbfsl{S_G}$.
For the code corresponding to  $G'$, we use the bit-channels with the same index as the code corresponding to  $G$, for transmission of information bits, and leave the rest as frozen. 
The probability of block error under SC decoding, which is described in the last part of this proof, for the code corresponding to  $G'$, $P_{e, G'}$, can be bounded above, as in \cite{arikan2009channel}, by the sum of the Bhattacharyya parameters of the bit-channels for the source bits (that are not frozen), that is,
$$
P_{e, G'} \leq 
\sum_{\mathbf{s}\in \mathbfsl{S_{G}}}Z_{G'}^{\mathbf{s}} 
\leq 
\sum_{\mathbf{s}\in \mathbfsl{S_{G}}} 2^{-N^\beta}
=
\abs{\mathbfsl{S_G} } 2^{-N^\beta},
$$ where the second inequality follows because, for  $\mathbf{s}\in  \mathbfsl{S_G} $, we must have  $\mathbf{s}\in  \mathbfsl{S_{G'}}$ and thus $Z_{G'}^{\mathbf{s}}\leq 2^{-N^\beta}$.
From \eqref{eq:rate_to_C}, for all sufficiently large $n$, we have 
     $P_{e, G'} \leq NC 2^{-N^\beta}.$
With some calculus, one may show that, for any $\beta'< \frac{1}{2}$, 
     $P_{e, G'} \leq 2^{-N^{\beta'}} $ for all sufficiently large $n$.

The second claim follows from the fact that the GM for the code corresponding to  $G'$ is a submatrix of $G'$, and the Hamming weight of each column of $G'$ is upper bounded by $w_{u.b.} = 2^{n\lambda}= N^\lambda$. 

The third claim is a {consequence} of Proposition \ref{prop:DRSrateloss} and Lemma \ref{lem:DRS_Z_BEC}. 
The number of information bits of the code corresponding to  $G'$ is given by 
$\abs{\mathbfsl{S_G} }$, and the length of the code is $N(1+\gamma)$. 
Hence the code rate is 
    $\frac{\abs{ \mathbfsl{S_G} }}{N(1+\gamma)}.$
Since the term $\gamma$ vanishes as $n$ grows large, we have 
\begin{equation}
\liminf_{n \to \infty} \frac{\abs{\mathbfsl{S_G}} }{N(1+\gamma)}
=
\liminf_{n \to \infty} \frac{\abs{\mathbfsl{S_G}} }{N}
= I(W) =C.    
\end{equation}

\begin{figure}[t]
     \centering
     \begin{subfigure}[b]{0.46\textwidth}
         \centering
         \includegraphics[trim=1.5cm 1.5cm 12.5cm 3.5cm, clip, width=\textwidth]{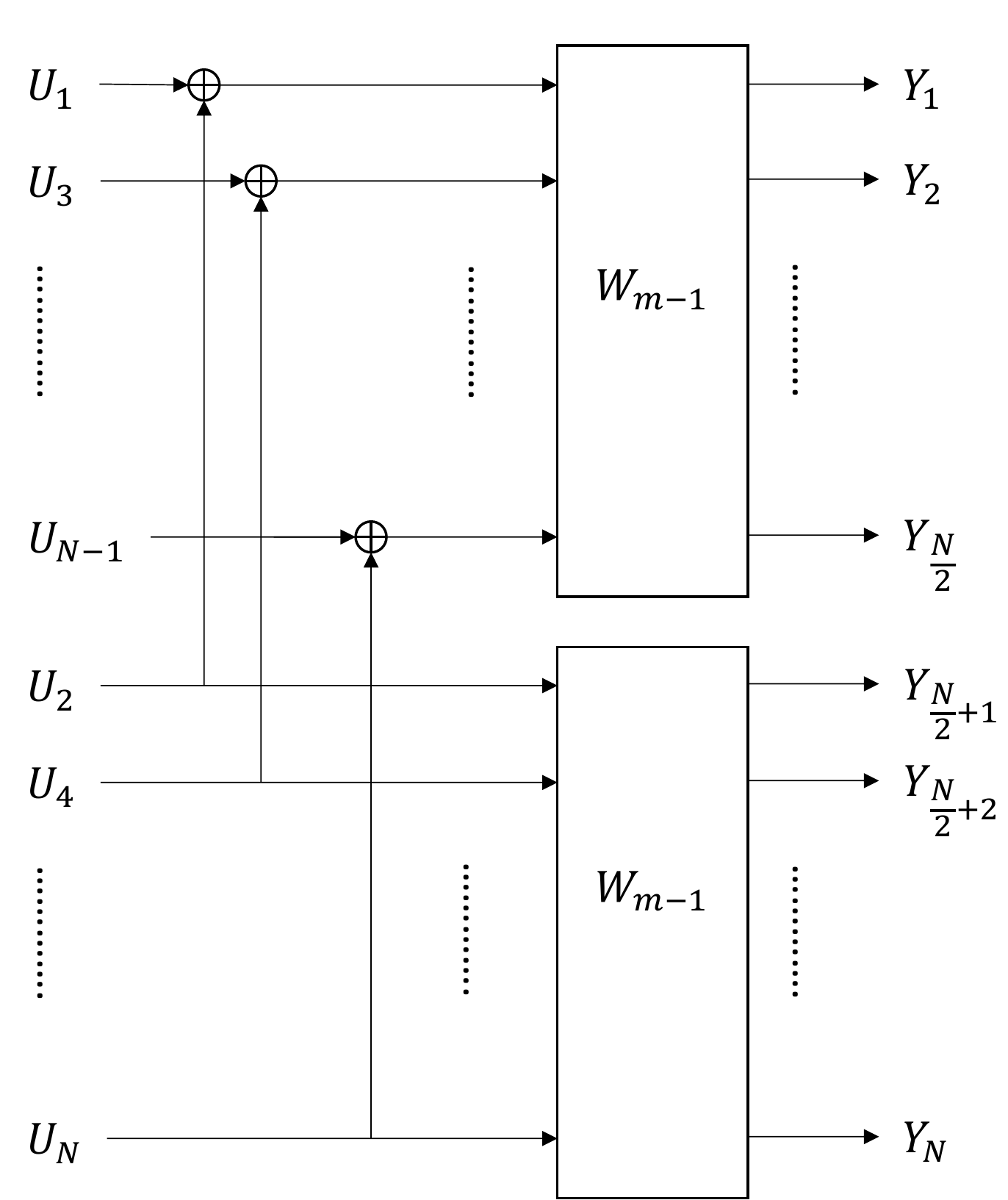}
         \caption{Encoding block for generator matrix $G_2^{\otimes m}$}
        \label{fig:G2m_BEC_EncoderWm}
     \end{subfigure}
     \hfill
     \begin{subfigure}[b]{0.48\textwidth}
         \centering
         \includegraphics[trim=2cm 1cm 10.5cm 2.5cm,  clip, width=\textwidth]{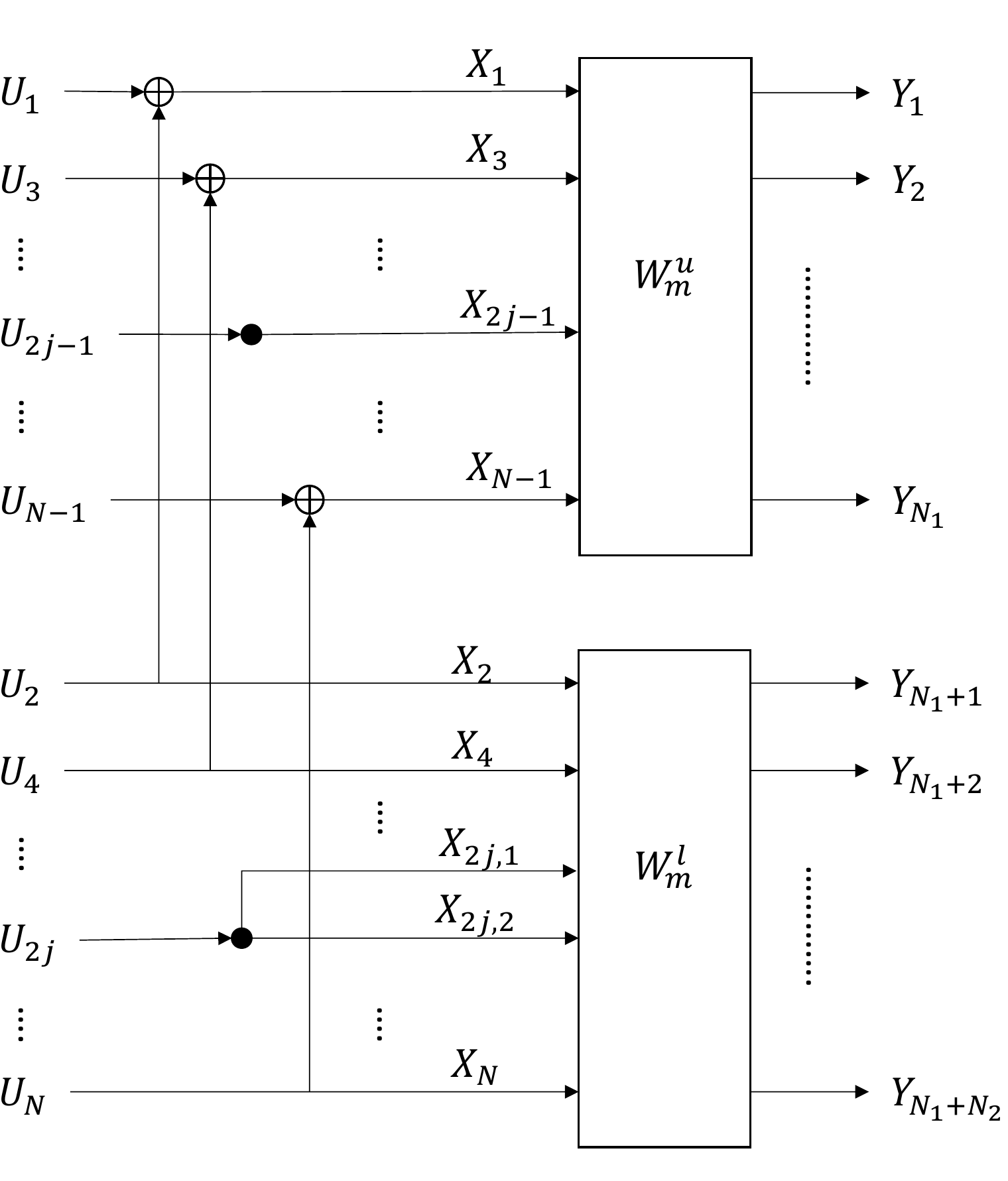}
        \caption{Encoding block for generator matrix $(G_2^{\otimes m})'$}
    \label{fig:G2m'_BEC_EncoderWm}
     \end{subfigure}
\caption{}
        \label{fig:Wm_BEC}
\vspace{-7mm}
\end{figure}

Finally, we prove the claim for the existence of a low-complexity decoder.
Just like that of the SC decoder for conventional polar codes with kernel $G_2$, the decoding algorithm proceeds in a recursive manner.
Let $U_1, \ldots, U_N$ be the inputs, and $Y_1, \ldots, Y_{N_1}, Y_{N_1 +1}, \ldots,  Y_{N_1+ N_2} $ the outputs, where $N_1+N_2 = N(1+\gamma)$, as shown in Figure\,\ref{fig:G2m'_BEC_EncoderWm}.
However, while the polar code based on $G_2^{\otimes n}$, as shown in Figure\,\ref{fig:G2m_BEC_EncoderWm}, is recursive in the encoder structure, i.e., the two encoding sub-blocks corresponding to $G_2^{\otimes n-1}$ are identical, the code based on $(G_2^{\otimes n})'$ is not, as the blocks $W_n^u$ and  $W_n^l$ are not necessarily equal. In fact, when there is a split in the GM due to the DRS algorithm, i.e., when one or more of the XOR operations shown in Figure\,\ref{fig:G2m'_BEC_EncoderWm} is replaced by two solid black circle, the number of inputs of the block $W_n^l$ will be larger than that of $W_n^u$. 

Let $\cF \subseteq \mathset{1,\ldots, N}$ be the set of the indices of the frozen bits. 
The decoder declares estimates $\hat{U}_i$ of the inputs, for $1\leq i \leq N$, sequentially by:
\begin{equation}\label{eq:UiestimateBEC}
    \hat{U_i} = \begin{cases}
u_i, &\mbox{ if } i\in \cF, \\
\psi_i(Y_1^{ N_1+N_2}, \hat{U}_1^{i-1} ,W_n) &\mbox{ if } i \notin \cF,
\end{cases}
\end{equation}
where     $\psi_i(Y_1^{ N_1+N_2}, \hat{U}_1^{i-1} ,W_n)$ can be found in following four cases, and $W_n$ denotes the encoding block shown in Figure \ref{fig:G2m'_BEC_EncoderWm}. 
Let the symbol $e$ denotes an erasure, and assume $e \oplus b = e$ for $b \in \mathset{0,1,e}$.

\begin{itemize}
    \item If $i$ is odd and $X_i = U_i \oplus U_{i+1}$, which corresponds to an unsplit XOR operation observed by $U_i$, \begin{equation}\
        \psi_i(Y_1^{ N_1+N_2}, \hat{U}_1^{i-1} ,W_n) \triangleq
    \begin{cases}
        \hat{X_i}\oplus \hat{X}_{i+1} &\mbox{ if } \hat{X_i}\neq e, \hat{X}_{i+1}\neq e, \\
        e, &\mbox{ otherwise.} 
    \end{cases}
    \end{equation}   

    \item If $i$ is odd and $X_i = U_i$, which corresponds to a split XOR operation, 
    $\psi_i(Y_1^{ N_1+N_2}, \hat{U}_1^{i-1} , W_n) \triangleq \hat{X}_i$.
    
    \item If $i$ is even and $X_{i-1} = U_i \oplus U_{i-1}$, which corresponds to an unsplit XOR operation,
    \begin{equation}
        \psi_i(Y_1^{ N_1+N_2}, \hat{U}_1^{i-1} , W_n ) \triangleq
    \begin{cases}
        \hat{X_i}, &\mbox{ if } \hat{X_i}\neq e, \hat{X}_{i-1} = e, \\&\mbox{ or if } \hat{X_i}\neq e, \hat{X}_{i-1} \neq e, \hat{X_i} = \hat{X}_{i-1} \oplus \hat{U}_{i-1},\\
         \hat{X}_{i-1} \oplus \hat{U}_{i-1}, &\mbox{ if } \hat{X_i}= e, \hat{X}_{i-1} \neq e, \hat{U}_{i-1} \neq e\\
        e, &\mbox{ otherwise.} 
    \end{cases}
    \end{equation}  
    
    \item If $i$ is even and $X_{i,1}=X_{i,2} = U_i$, which corresponds to a split XOR operation,
    \begin{equation}
        \psi_i(Y_1^{ N_1+N_2}, \hat{U}_1^{i-1} , W_n ) \triangleq
    \begin{cases}
        \hat{X}_{i,1}, &\mbox{ if } \hat{X}_{i,1}\neq e, \hat{X}_{i,2} = e, \mbox{ or if } \hat{X}_{i,1} = \hat{X}_{i,2} \neq e,\\
        \hat{X}_{i,2}, &\mbox{ if } \hat{X}_{i,1}= e, \hat{X}_{i,2} \neq e,\\
        e, &\mbox{ otherwise.} 
    \end{cases}
    \end{equation} 
\end{itemize}
The estimates $\hat{X}_1,\hat{X}_3,\ldots, \hat{X}_{N-1}$ and $\hat{X}_{2}, \hat{X}_{4}, \ldots, \hat{X}_{2j,1} , \hat{X}_{2j,2}, \ldots, \hat{X}_{N}$ are found in a similar approach using the blocks $W_n^u$ and $ W_n^l$ along with the outputs $Y_1, \ldots, Y_{N_1}$ and  $Y_{N_1 +1}, \ldots,  Y_{N_1+ N_2} $, respectively. 

For the right-most variables, the blocks they observe are identical copies of the BEC $W$.  Hence the estimates of the variables, denoted as $\hat{X}_1^{(n)}, \hat{X}_2^{(n)}, \ldots, \hat{X}_{N_1+ N_2}^{(n)}$ are naturally defined by the outputs of the channels, i.e., $\hat{X}_i^{(n)} = Y_i$ for $i = 1, 2, \ldots , N_1+N_2.$

At each stage there are at most $ N_1+N_2= N(1+\gamma) = O(N)$ estimates to make, and the recursion ends in $\log(N)$ steps.
Since each estimate is obtained with constant complexity, the total decoding complexity for the code based on \textproc{DRS}($G_2^{\otimes n}$) is bounded by $O(N\log N)$.
\endproof

\begin{figure}[h!]
     \centering
     \begin{subfigure}[b]{0.49\textwidth}
         \centering
         \includegraphics[width=\textwidth]{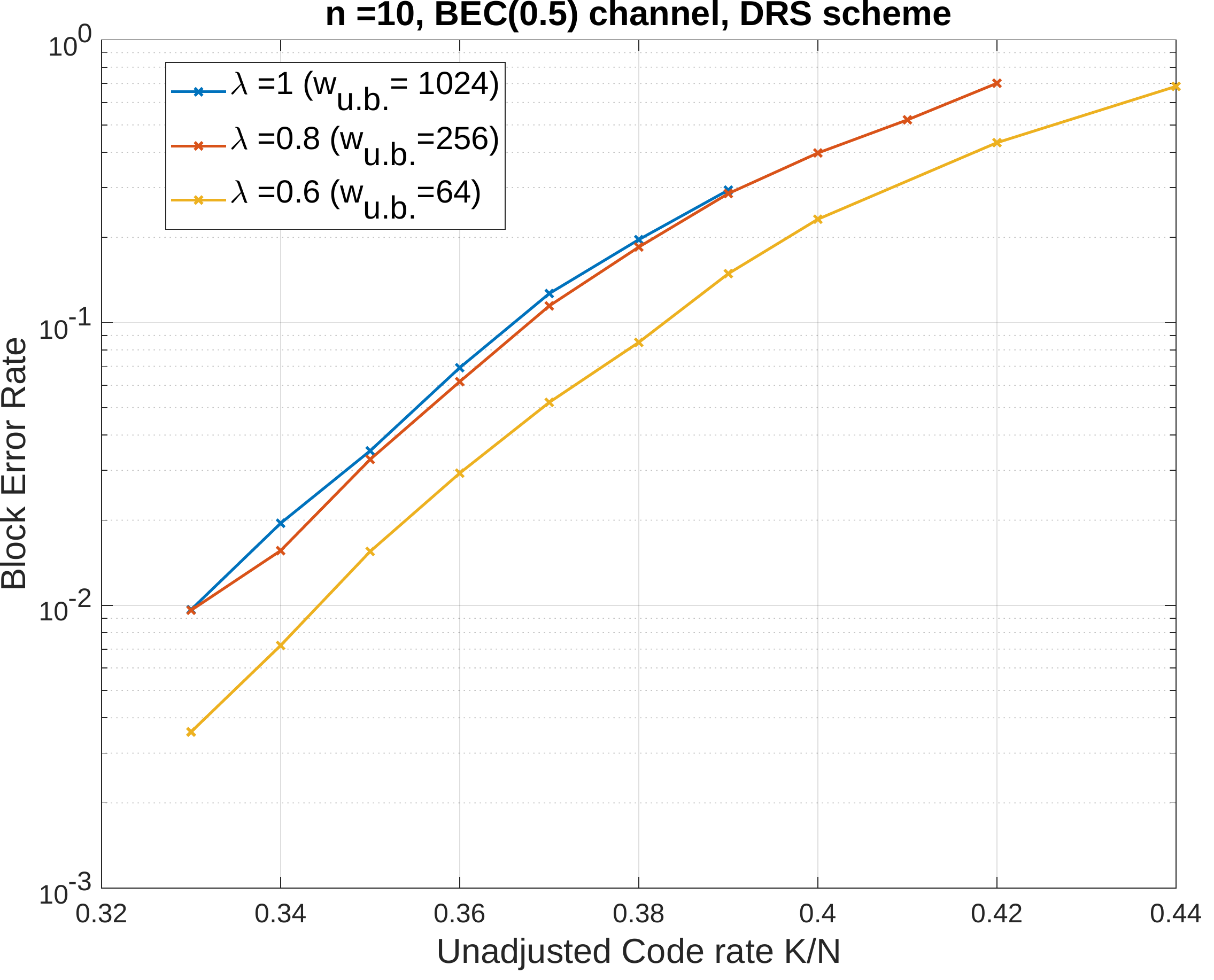}
         \caption{Unadjusted code rate}
        \label{fig:n10_Pe_DRM_unadj}
     \end{subfigure}
     \hfill
     \begin{subfigure}[b]{0.49\textwidth}
         \centering
         \includegraphics[width=\textwidth]{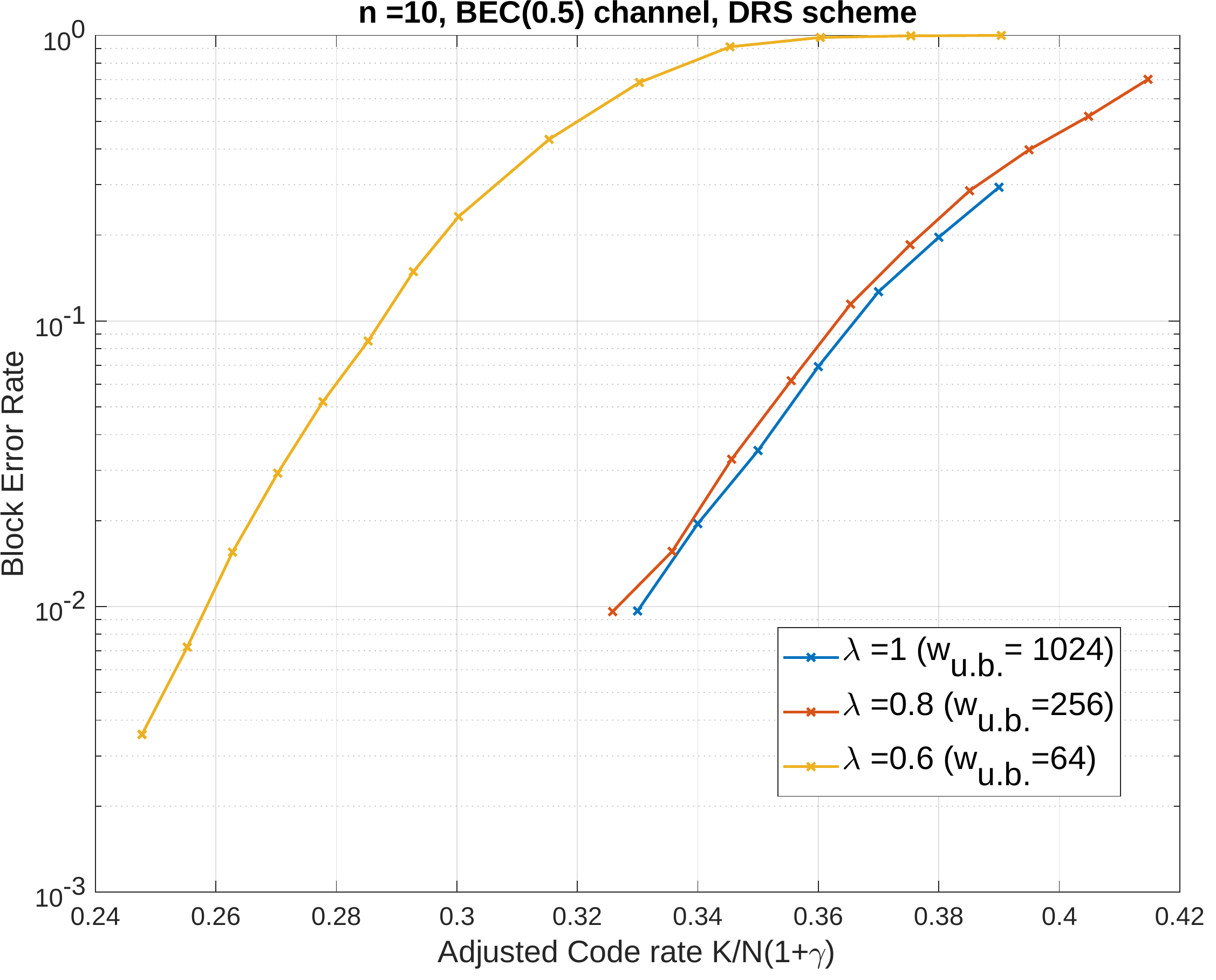}
        \caption{Adjusted code rate}
    \label{fig:n10_Pe_DRM_adj}
     \end{subfigure}
\caption{Error probability for polar-DRS codes with $n=10$ with $w_{u.b.} \in\mathset{64, 256, 1024}$}
        \label{fig:n10_Pe_DRS}
\vspace{-5mm}
\end{figure}

We evaluate the performance of the polar-DRS codes with $n =10$ and $\lambda =0.6, 0.8, 1.0$, under the SC decoding scheme described in the proof of Theorem \ref{Thm:BEClowcomplDEC}, over the BEC with erasure probability $\epsilon = 0.5$. 
In Figure~\ref{fig:n10_Pe_DRM_unadj}, the block error probabilities for the curves with smaller $\lambda$ are smaller, due to the improvement of some of the Bhattacharyya parameters observed by the information bits. That is, there are sequences $(s_1, s_2, \ldots, s_n)\in \mathset{-,+}^n$ for which the inequality in Lemma~\ref{lem:DRS_Z_BEC} is strict. However, after factoring in multiplicative rate loss $\gamma$, we may observe in Figure~\ref{fig:n10_Pe_DRM_adj} that the performance of the codes with $\lambda = 0.6$ are substantially worse than the original polar code ($\lambda =1$ curve), and those with $\lambda = 0.8$ deliver trade-off between code rate and error probability comparable to the original polar code, while guaranteeing the threshold $w_{u.b.}$ is one-fourth of the latter.

\begin{remark} \label{rem:DRSnotForBMS} 
We note that for general BMS channels, Lemma \ref{lem:DRS_Z_BEC} may fail. One key part in the proof (see Appendix \ref{appendix:proof_LCdec_BEC}) 
is the fact that the 
Bhattacharyya parameter for the bit-channel observed by $U_i$ is a non-decreasing function of those of $W(X_1), \ldots, W(X_{f{(m)}})$ for  $i \leq 2^m$, and of $W(X_{f{(m)}+1}), \ldots,$  $W(X_{2f{(m)}})$ for  $i >2^m$, when all the channels are BECs. 
We now provide an example where we see the argument for Lemma \ref{lem:DRS_Z_BEC} fail for BMS channels. 
Let $a, a', b, b'$ be four distinct elements and $\mathcal{Y} = \mathset{a, a', b, b'}$. 
Let two BMS channels $W_1, W_2: \mathset{0, 1} \rightarrow \mathcal{Y}$ be given, and that  $W_1( y | 0) = W_1(\phi(y) |1)$ and $W_2(y | 0) = W_2( \phi(y) |1)$ for all $y \in \mathcal{Y}$ where the involution $\phi$ maps $a \mapsto a', b \mapsto b'$. 
Assume the channel transition probabilities are $W_1(a | 0)=    6/9, W_1(b | 0)=   1/9 $, $W_1(b' | 0)=     1/9, W_1(a' | 0)=     1/9$ and $ W_2(a | 0)=    5/11$, $W_2(b | 0) = 4/11$, $W_2(b' | 0)=     1/11$, $W_2(a' | 0)=1/11$. The Bhattacharyya parameters for $W_1, W_2$ are    respectively $0.7666 $     and $0.7702$. If $m=1$ and $B_m$ is simply the kernel $G_2$, the symbols $X_1, X_2$ are functions of $U_1, U_2$ given by 
$X_1 = U_1 +U_2$ and $X_2 = U_2$. 

We now consider two possible cases for the pair $(W(X_1), W(X_2)) $. 
If $(W(X_1), W(X_2)) = (W_1, W_2 )$, the Bhattacharyya parameters for the bit-channels observed by $U_1, U_2$ are respectively  $    0.9147$ and   $0.5904$. 
If $(W(X_1), W(X_2)) = (W_2, W_2 )$, the Bhattacharyya parameters for the bit-channels observed by $U_1, U_2$ are respectively  $    0.9137$ and   $0.5932$. 
We note that while the Bhattacharyya parameters for $W(X_1), W(X_2)$ in the second case are no less than in the first case, the Bhattacharyya parameter $Z(U_1)$ in the second case is smaller than in the first case. 
With the above observation, one can not claim the validity of Theorem \ref{Thm:BEClowcomplDEC} for general BMS channels.  
This motivates a new code construction for general BMS channels.
\end{remark}

\subsection{Low-complexity Decoder for Polar-based Codes: BMS}
\label{subsection:LCdec_BMS}
This section introduces a capacity-achieving polar-based coding scheme with low-complexity decoder for general BMS channels. 
For general BMS channels, the Bhattacharyya parameter of the bit-channel $W^-$ cannot be expressed only in terms of parameters of the channel $W$. This implies that Lemma \ref{lem:DRS_Z_BEC} and Theorem \ref{Thm:BEClowcomplDEC} are not applicable for channels other than BEC, as pointed out in Remark \ref{rem:DRSnotForBMS}. 
A procedure that augments the generator matrix corresponding to  $G'$, the output of the DRS algorithm for the matrix $G_2^{\otimes n}$, may be used to construct a capacity-achieving linear code over any BMS channel $W$.

\subsubsection{ADRS Scheme}\label{subsub:ADRS}

The encoding scheme, termed augmented-DRS (ADRS) scheme, avoids heavy columns in the GM and, at the same time, guarantees that the bit-channels observed by the source bits $U_i$ have the same statistical characteristics as when they are encoded with the generator matrix $G_2^{\otimes n}$.
Specifically, the ADRS scheme modifies the encoder for $G_2^{\otimes n}$ starting from the split XOR operations associated with the first polarization recursion, then the second recursion, and proceed all the way to the $n$-th recursion, where a XOR operation is split if and only if it is split in an encoder with generator matrix \textproc{DRS}($G_2^{\otimes n}$).

\begin{figure}
      \centering
      \begin{subfigure}[b]{0.45\textwidth}
          \centering
		\includegraphics[trim=2cm 2.5cm 15cm 15cm, clip,  width= \textwidth]{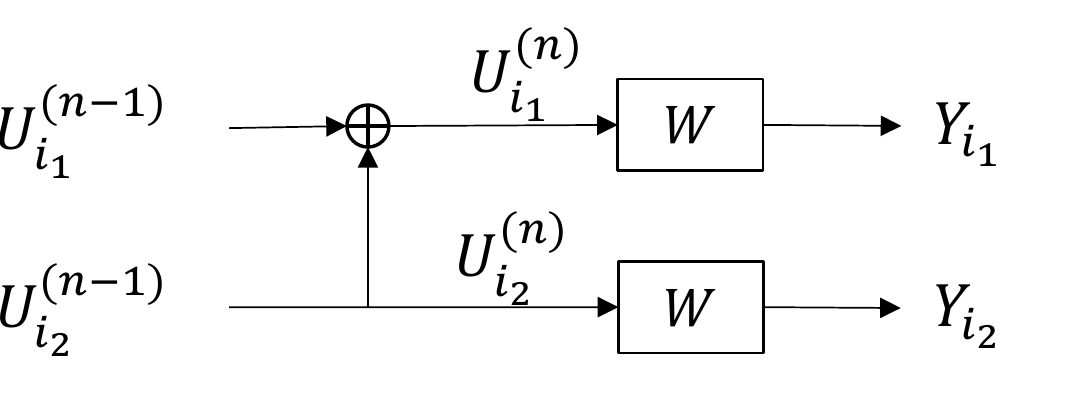}
          \caption{Before Modification}
          \label{fig:A-DRS-generalUnsplit_j1}
      \end{subfigure}
      \begin{subfigure}[b]{0.4\textwidth}
          \centering
		\includegraphics[trim=2cm 1cm 14cm 14cm, clip,  width= \textwidth]{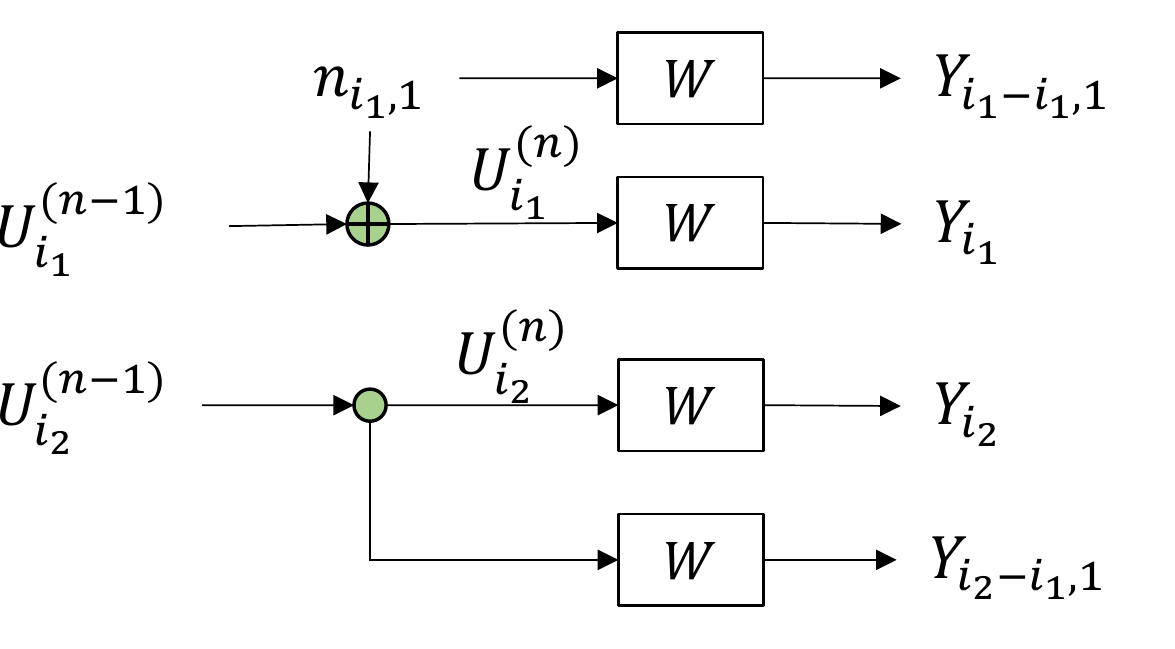}
          \caption{ADRS modification}
          \label{fig:A-DRS-generalSplit_j1}
      \end{subfigure}
     \caption{ADRS scheme for a split XOR of first iteration of polarization}
     \label{fig:A-DRS-j=1}
\vspace{-7mm}
\end{figure}

Assume an XOR operation with operands $U_{i_1}^{(n-j)}$ and $U_{i_2}^{(n-j)}$ and the output $U_{i_1}^{(n-j+1)}$, where $i_1 = \mathsf{Bi2De}(s_1, s_2, \ldots,s_{j-1}, s_j =-, s_{j+1}, \ldots, s_{n}  )+1$ and $i_2 = \mathsf{Bi2De}(s_1, s_2, \ldots, s_{j-1}, s_j =+, s_{j+1}, \ldots, s_{n} )+1= i_1 +2^{n-j}$, is to be split (see Section \ref{subsection:LCdec_BEC} for the function $\mathsf{Bi2De}(\cdot)$). 
If $j =1$, before modification, the variables $U_{i_1}^{(n)}$ and $U_{i_2}^{(n)}$ are transmitted through two copies of $W$, and the bit-channels observed by $U_{i_1}^{(n-1)}$ and $U_{i_2}^{(n-1)}$ are $W^{-}$ and $W^{+}$, respectively, as shown in Figure\,\ref{fig:A-DRS-generalUnsplit_j1}. 
If the XOR operation is split according to \textproc{DRS}($G_2^{\otimes n}$), ADRS scheme replaces the structure by that given in Figure\,\ref{fig:A-DRS-generalSplit_j1}, where $n_{i_1, 1}$ is a Bernoulli($0.5$) random variable independent of all the other variables. 

If $j \geq 2$, assume that the ADRS modification for the split operations for the first $(j-1)$ recursions are completed. 
Let $n_{i_1, j}$ be a Bernoulli($0.5$) random variable independent of all the other given variables. 
The part of encoding diagram to the right of $U_{i_1}^{(n-j+1)}$ is replicated, where  $n_{i_1, j}$ takes the place of $U_{i_1}^{(n-j+1)}$ in the replica. And then we let  $U_{i_1}^{(n-j+1)} = U_{i_1}^{(n-j)}\oplus n_{i_1, j}$. 
In addition, the part of encoding diagram to the right of $U_{i_2}^{(n-j+1)}$ is replicated, 
and a copy of $U_{i_2}^{(n-j)}$ is transmitted through the replica. The variable  $U_{i_2}^{(n-j+1)} $ remains  $U_{i_2}^{(n-j+1)}= U_{i_2}^{(n-j)}$. 
 
\begin{figure}[t]
     \centering
     \begin{subfigure}[b]{0.46\textwidth}
	    \centering
    	\includegraphics[trim=1.5cm 1.5cm 9.5cm 8.5cm, clip,  width= \textwidth]{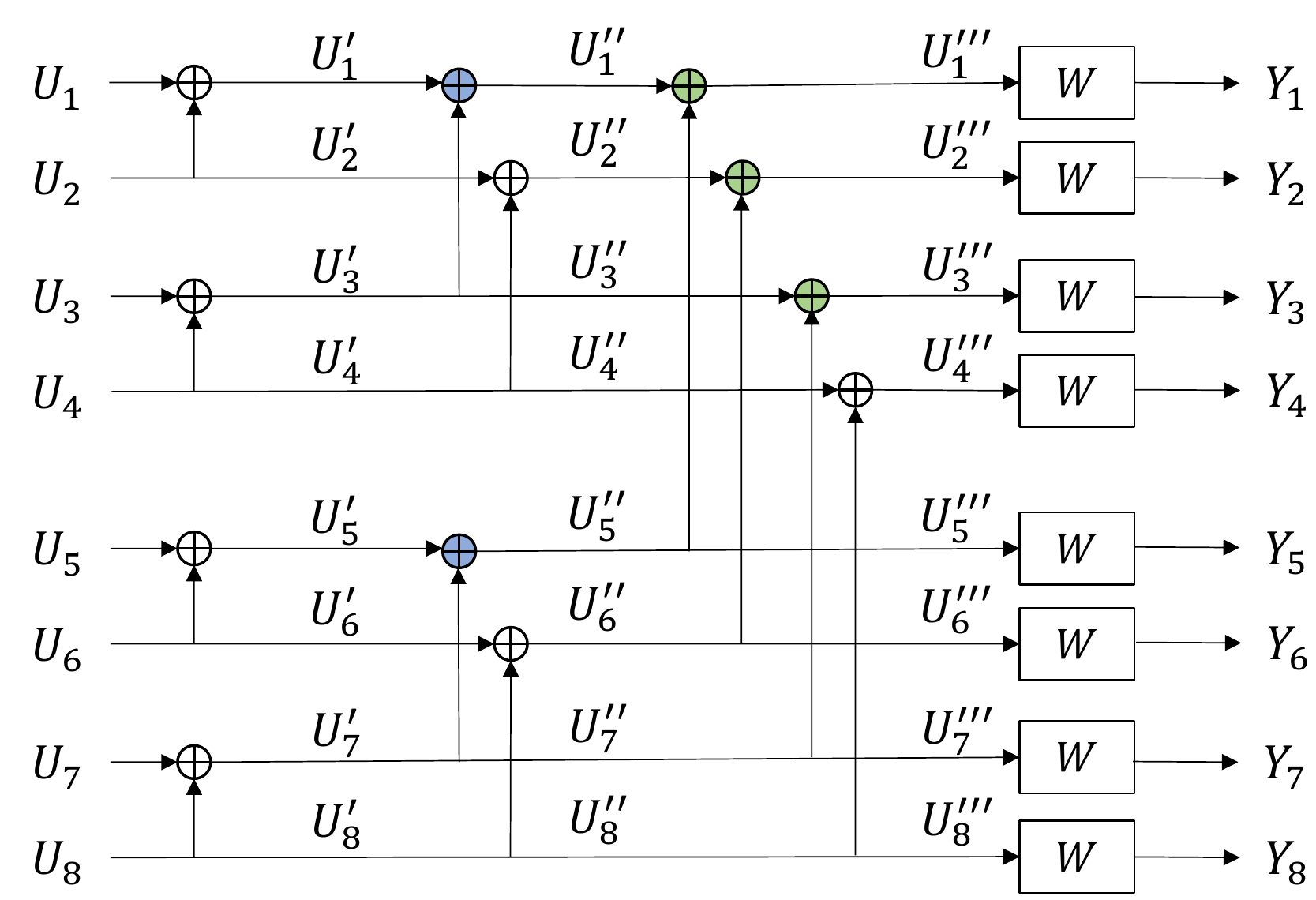}
    	\caption{Encoding diagram for $G_2^{\otimes 3}$}\label{fig:ADRS_unsplit}
     \end{subfigure}
     \hfill
     \begin{subfigure}[b]{0.48\textwidth}
    	\centering
    	\includegraphics[trim=1cm 1.5cm 5cm 2.5cm, clip, width= \textwidth]{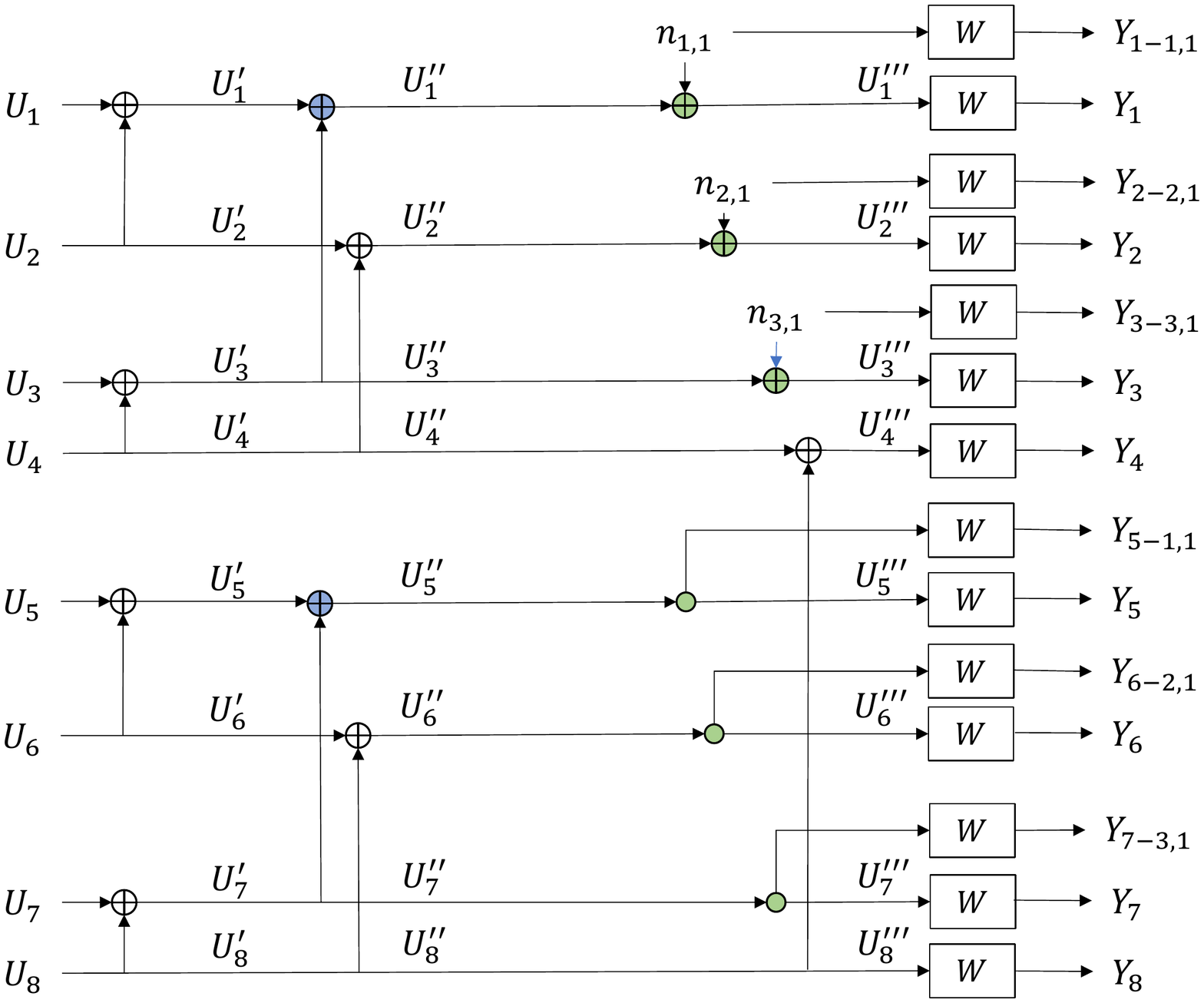}
    	\caption{ADRS for splits corresponding to $s_1$ in $G_2^{\otimes 3}$ }\label{fig:ADRS_1st_iteration}
     \end{subfigure}\vspace{2mm}
     \begin{subfigure}[b]{\textwidth}
        \centering
        \includegraphics[trim=4.7cm 22.6cm 2.4cm 0.3cm, width= \textwidth]{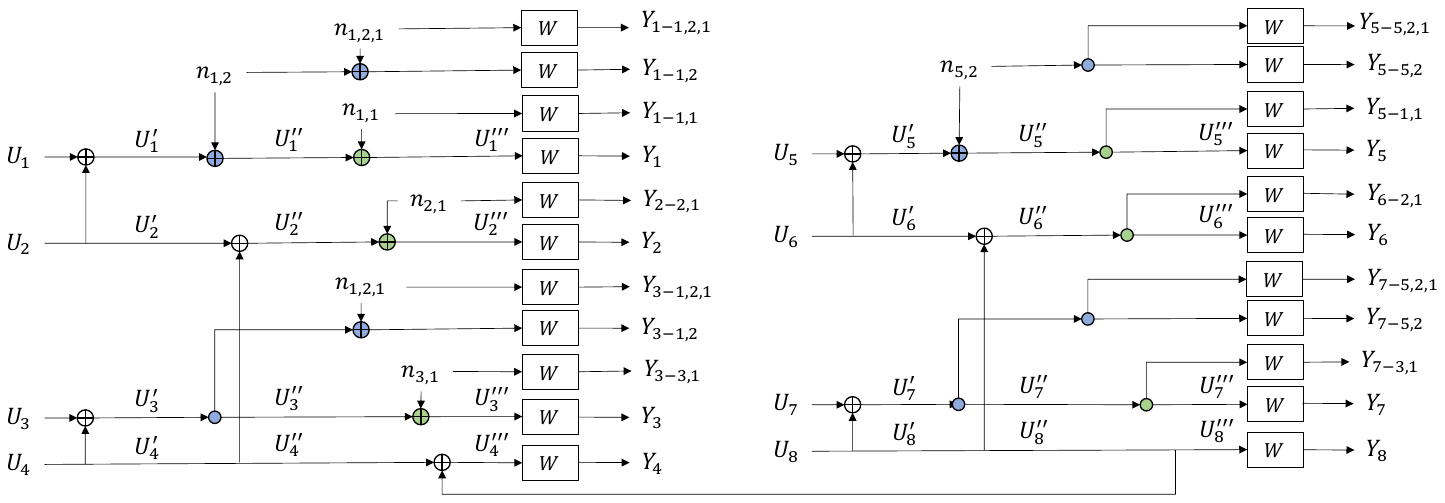}
	    \caption{ADRS Encoding Diagram for $G_2^{\otimes 3}$ with $w_{u.b.}=2$}\label{fig:ADRS_n3_entire}
     \end{subfigure}
\caption{ADRS example with $N=8$ and $w_{u.b.}=2$}
        \label{fig:Wm_BEC}
\vspace{-7mm}
\end{figure}

We demonstrate the procedure described above through the following example. 
Assume $n=3$, $N=8$, and $w_{u.b.} =2$. 
The encoding diagram for $G_2^{\otimes 3}$ is shown in Figure\,\ref{fig:ADRS_unsplit}, and the XOR operations that are split in \textproc{DRS}($G_2^{\otimes 3}$) are marked in green and blue, which indicate the operations are due to the first and the second polarization recursions (i.e., $s_1$ and $s_2$), respectively. 
The notations $U_i', U_i'', U_i'''$ are used to represent $U_i^{(1)}, U_i^{(2)}, U_i^{(3)}$. 
Replacing the XOR operations marked in green as described for the case of $j=1$, the encoding diagram is now shown in Figure\,\ref{fig:ADRS_1st_iteration}. 
For the XOR operations marked in blue, we proceed by using the step for $j\geq 2$ and obtain the diagram shown in Figure\,\ref{fig:ADRS_n3_entire}. 

It can be noted that the bit-channels observed by each of   $U_i^{(j)}$, for $i= 1, 2, \ldots, N$ and $j = 0,1,2, \ldots, n$, in the ADRS encoder are the same as those in the standard encoder for the generator matrix $G_2^{\otimes n}$ (The variable $U_i^{(0)}$ are given by $U_i$ for $1\leq i \leq N$). 
When an XOR operation associated with the $j$-th recursion, with operands $U_{i_1}^{(n-j)}$ and $U_{i_2}^{(n-j)}$ and the output $U_{i_1}^{(n-j+1)}$, is split and modified under the ADRS scheme, the  complexity of computing the likelihood or log-likelihood for $U_{i_1}^{(n-j)}$ and $U_{i_2}^{(n-j)}$ can be upper bounded by $2(2^1 + 2^2+ \ldots + 2^j)c = 2(2^{j+1} -2)c$, for some constant $c>0$.

\subsubsection{Polar-ADRS Code Performance}
We are now ready to show the performance of the polar-based code whose encoding structure is given by the ADRS scheme, referred to as the \textit{polar-ADRS} code. First we show the existence of a low-complexity decoder. 
\begin{proposition}\label{prop:ADRScompl}
    Let a constant $\lambda > \lambda^{\dagger} \triangleq (\log_2 3)^{-1} \approx 0.631$ be given.
    The decoding complexity for a SC decoder for the polar-ADRS code is bounded by $\bigoh(N\log N)$ for all sufficiently large $n$ if the threshold for the DRS algorithm is $w_{u.b.} = 2^{n\lambda}$. 
\end{proposition}
\proof The proof is provided in Appendix \ref{pf:prop:ADRScompl}. \endproof

Second, it can be observed that the number of additional copies of channels due to the modification for an XOR operation at the $j$-th polarization recursion is $2^j$. 
We find the total number of extra channel uses and the ratio $\gamma$ of that to the number $N= 2^n$ of channel uses for the code corresponding to $G_2^{\otimes n}$ in the following.  Assume that the column weight threshold of the DRS algorithm is given by $w_{u.b.} = 2^{n\lambda}$.

\begin{proposition}\label{Prop:BMSrateloss}
    Let $N(1+\gamma)$ be the number of channel uses of the encoder for the ADRS scheme based on \textproc{DRS}($G_2^{\otimes n}$) with $w_{u.b.} = 2^{n\lambda}$.
    Then the term  $\gamma$ goes to $0$ as $n$ grows large, if we have $\lambda > \lambda^{\dagger}$.
\end{proposition}
\proof The proof is provided in Appendix \ref{pf:Prop:BMSrateloss}. \endproof

We are ready to show the existence of a sequence of capacity-achieving codes over general BMS channels with GMs where the column weights are bounded by a polynomial in the blocklength, and that the block error probability under a low  complexity decoder vanishes as the $n$ grows large. 
Note that while this result is also applicable when the underlying channel is a BEC, the constraint on $\lambda$ is stricter than that in Theorem~\ref{Thm:BEClowcomplDEC}, due to the difference in the encoding and decoding schemes. 
\begin{theorem}    \label{Thm:BMSlowcomplDEC}
    Let $\beta < E(G_2)=0.5 $, $\lambda > \lambda^{\dagger}$, and a BMS channel $W$ with capacity $C$ be given. There exists a sequence of codes 
    with the following properties for all sufficiently large $n$:
        (1) The error probability under SC decoding is upper bounded by $2^{-N^{\beta}}$, where $N= 2^n$,
        (2) The Hamming weight of each column of the GM is upper bounded by $N^\lambda$,
        (3) The rate approaches $C$ as $n$ grows large, and
        (4) The codes can be decoded by a SC decoding scheme with complexity $\bigoh(N\log N)$.
    
\end{theorem}
\proof 
We prove the four properties in order as follows.
First, similar to the proof of Theorem\,\ref{Thm:BEClowcomplDEC}, for $i =1, 2,\ldots, N$, the bit $U_i$ is frozen in the polar-ADRS code with rate $R< C$ if and only if it is frozen in the polar code with kernel $G_2$, blocklength $N= 2^n$, and the rate $R$.
Hence, the probability of error of the polar-ADRS code can be bounded in the same way as its polar-code counterpart, since the bit-channels observed by the source bits $U_i$, and the corresponding Bhattacharyya parameters, are identical to those when they are encoded with the standard polar code.

Second, when the ADRS scheme is based on \textproc{DRS}($G_2^{\otimes n}$) with $w_{u.b.} = 2^{n\lambda}$, 
the generator matrix for the polar-ADRS code is a submatrix of \textproc{DRS}($G_2^{\otimes n}$). 
The column weights of the GM for the polar-ADRS code are thus upper bounded by $w_{u.b.} = 2^{n\lambda} = N^{\lambda}$.
The third claim holds by using an argument similar to the one used in the proof of Theorem\,\ref{Thm:BEClowcomplDEC}. This is because the term $\gamma$ vanishes as $n$ grows large according to Proposition\,\ref{Prop:BMSrateloss}.
Finally, note that the fourth claim is equivalent to Proposition\,\ref{prop:ADRScompl}.
\endproof

\section{Conclusion}\label{sec:Conclusion} 
This paper provided three constructions for capacity-achieving linear codes, based on polar coding, where all the GM column weights are upper bounded sublinearly in the block length. 
The first construction is a sequence of polar codes based on general polarization kernels where the GM column weights are upper bounded by $N^s$ for any fixed $s>0$, and allows the codes to be decoded by a SC decoder.
In order to attain a better trade-off between the GM sparsity and the fall in error probability, we then proposed a column-splitting algorithm for the GM, termed the DRS algorithm.
With the DRS algorithm, we designed two encoding schemes which yield two polar-based codes, referred to as polar-DRS codes and polar-ADRS codes, that are decodable with low-complexity decoders for the BECs and general BMS channels, respectively.
The polar-based codes preserve several fundamental properties of the standard polar code with $G_2$ kernel including the asymptotic error rate upper bound and decoding complexity. 
Further, the GM column weights of the polar-DRS  and polar-ADRS codes are bounded from above by $N^\lambda$, for $\lambda\approx 0.585$ and $\lambda\approx 0.631$, respectively, while the best bound for the standard polar codes scales linearly in $N$.
The proposed constructions are also distinct from known constructions for codes with constraints on the GM sparsity by having analytical error probability upper bounds scaling as $\bigoh(2^{-N^{t}})$ under SC decoders.
A future direction is to design splitting algorithm and/or encoding schemes that  preserve key properties of the polar codes based on general polarization kernels, and show that the corresponding  polar-based codes exhibit better sparsity versus error rate trade-off. 

\appendix
\label{sec:Appendix}

\subsection{Proof for Section \ref{sec:SparseLDGM}}\label{pf:sec:SparseLDGM}

\textit{{Proof of Proposition \ref{prop:noLogColumns}}: }\label{pf:prop:noLogColumns}
    Since $G_l$ is a polarization kernel, there is at least one column in $G_l$ with weight at least $2$. 
    To see this, note that $G_l$ being invertible implies that all rows and columns are nonzero vectors. Now, if all the columns of $G_l$ have weight equal to $1$, then all the rows must also have weight equal to $1$, i.e., $G_l$ is a permutation matrix. Then $D_i =1, \forall i$, and $ E(G_l) =0$, which implies that $G_l$ can not be polarization kernel. The contradiction shows that at least one column in $G_l$ must have a weight at least $2$. 
    
    Let $k \geq 1$ denote the number of columns in $G_l$ with a weight at least $2$. 
    Let \textbf{v} be a randomly uniformly chosen column of $G_l^{\otimes n}$, and $w(\textbf{v}) $ be the Hamming weight of \textbf{v}. 
    For $\frac{1}{l} > r > 0$,
    \begin{align*}
        &\Pr\left( w(\textbf{v}) =  O(N^{r \log_l2}) \right)
        \leq \Pr \left(2^{\sum_{i=1}^n F_i}   = O(N^{r \log_l2}) = O(2^{nr} ) \right), 
    \end{align*}
    where $F_i $ is the indicator variable that one of $k$ non-unit-weight columns is used in the $i$-th Kronecker product of $G_l$ to form \textbf{v}. 
    The variables $F_1, F_2, \ldots, F_n$ are i.i.d. as $\text{Ber}(k/l) $. 
    Law of large numbers implies that
    $\sum_{i=1}^n F_i \geq \frac{kn}{l} > nr$ with high probability.
    Thus, $ \Pr \left(2^{\sum_{i=1}^n F_i} = O( 2^{nr} ) \right) \to 0$  as $n \to \infty$ for any $r>0$.
\endproof

\subsection{Proofs for Subsection \ref{sub:DRSalg}}\label{appendix:proof_DRSalg}
\subsubsection{ }\label{pf:lem:rateLossIndepOfOrder}
In order to understand the DRS algorithm's effect on $G= G_2^{\otimes n} $, we first study how the order of two special Kronecker product operations affects the number of output vectors.
We present the following Lemmas \ref{lem:LR=RL} and \ref{lem:DRSforTwoVectors} toward the proof of Lemma \ref{lem:rateLossIndepOfOrder}.

\begin{lemma} \label{lem:LR=RL}
Let a column vector $v$ and a column weight threshold $w_{u.b.}$ be given. Then the outputs of the DRS algorithm for
$(v \otimes     [1,1]^T )    \otimes [0,1]^T$
 and 
$(v \otimes     [0,1]^T )    \otimes  [1,1]^T$
contain the same number of vectors.
\end{lemma}
\proof{} 
The input vectors can be denoted by:
\begin{align*}
(v \otimes     [1,1]^T)
    \otimes [0,1]^T
    \equiv v_{LR}
\; \mbox{ and } \;
(v \otimes     [0,1]^T )
    \otimes [1,1]^T
    \equiv v_{RL}.
\end{align*}    

We note that $ 2 w_H(v) =w_H(v_{LR}) = w_H(v_{RL}) $, and prove the lemma in two  cases:
\begin{enumerate}
    \item $2 w_H(v) \leq w_{u.b.}$:
        In this case, the algorithm will not split either $ v_{LR}$ or $v_{RL}$. Both outputs contain exactly one vector.
    
    \item $2 w_H(v) > w_{u.b.}$:
    Let $n_{DRS}(v)$ denote the size, or more precisely, the number of column vectors the DRS algorithm returns when it is applied to $v$.
    
    For $ v_{LR}$, the DRS algorithm observes $\mathbf{x}_h=\zero$, hence the number of output vectors is the same as the size of \textproc{DRS-Split}($w_{u.b.}, (v^T, v^T)^T $) (see Section \ref{sub:DRSalg}). 
    With $2 w_H(v) > w_{u.b.}$, the size of \textproc{DRS-Split}($(v^T, v^T)^T $) is the sum of the sizes of  $Y_h =$ \textproc{DRS-Split}($w_{u.b.}, \mathbf{x}_h=v$) and $Y_t =$ \textproc{DRS-Split}($w_{u.b.}, \mathbf{x}_t=v$). By assumption,  $\abs{Y_h}=\abs{Y_t}= n_{DRS}(v)$, giving     $n_{DRS}(v_{LR})=2\, n_{DRS}(v) $.

    For $ v_{RL}$, the number of vectors in the DRS algorithm output is the sum of the sizes of two sets $Y_h =$ \textproc{DRS-Split}($w_{u.b.}, \mathbf{x}_h=(\zero^T, v^T)^T $) and $Y_t =$ \textproc{DRS-Split}($w_{u.b.}, \mathbf{x}_t=(\zero^T, v^T)^T$).
    It is easy to see that $\abs{Y_h}=\abs{Y_t}= \abs{
     \textproc{DRS-Split}(w_{u.b.}, v)
    }$, which equals $n_{DRS}(v)$.
    Thus, $n_{DRS}(v_{RL})=2\, n_{DRS}(v) $.

\end{enumerate}
\endproof

The next lemma shows the effect of the DRS algorithm from a different perspective. If there are two vectors with the same column weights, and numbers of vectors of the DRS  algorithm outputs are identical when they are the inputs, the properties will be preserved when they undergo some basic Kronecker product operations. 

\begin{lemma}\label{lem:DRSforTwoVectors}
    Let $u_1$ and $u_2$ be two vectors with equal Hamming weights. Assume, for a given $w_{u.b.}$, the DRS algorithm splits $u_1$ and $u_2$ into the same number of vectors. Then the DRS algorithm also returns the same number of vectors for $u_1 \otimes [1,1]^T$ and $u_2\otimes [1,1]^T$, as well as for $u_1 \otimes [0,1]^T$ and $u_2\otimes [0,1]^T$. 
\end{lemma}
\proof
We first discuss the case when $u_1 \otimes [1,1]^T$ and $u_2\otimes [1,1]^T$ are processed by the DRS algorithm. 
If $2w_H(u_1)= 2w_H(u_2) \leq w_{u.b.}$, no splitting is done. 
If  $2w_H(u_1)= 2w_H(u_2) > w_{u.b.}$,  the size of the DRS algorithm output for the input $u_1 \otimes [1,1]^T$ is the sum of the sizes of $Y_h =$ \textproc{DRS-Split}($w_{u.b.}, \mathbf{x}_h=u_1$) and $Y_t =$ \textproc{DRS-Split}($w_{u.b.}, \mathbf{x}_t=u_1$), both of which are $n_{DRS}(u_1)$. 
The size of the output for the input $u_2 \otimes [1,1]^T$ can be found in a similar way to be $2 n_{DRS}(u_2)$.
Note that $n_{DRS}(u_1)=n_{DRS}(u_2)$ by assumption. Therefore, the sizes of the outputs of the DRS algorithm, when $u_1 \otimes [1,1]^T$ and $u_2\otimes [1,1]^T$ are the inputs, are equal.
Similarly, one can easily show that when $u_1 \otimes [0,1]^T$ and $u_2\otimes [0,1]^T$ are processed by the DRS algorithm, the number of output columns are equal.
\endproof
\vspace{2mm}

\textit{Proof of Lemma \ref{lem:rateLossIndepOfOrder}:}
Suppose that there is an index $i$ such that $(s_i, s_{i+1} )= (+,-)$.  
Let $v^{(i+1)}$ and $(v^{(i+1)})'$ be defined by (\ref{eq:vi_recursion}) with sequences $(s_1, \ldots, s_{i-1}, s_i= +, s_{i+1} = -)$ and $(s_1, \ldots, s_{i-1}, s'_i= -, s'_{i+1} = +)$, respectively.
We note that 
\begin{align*}
    v^{(i+1)} = \left( v^{(i-1)} \otimes     
    [0,1]^T
    \right) \otimes 
    [1, 1]^T
    \; \mbox{ and } \;
    (v^{(i+1)})' = \left( v^{(i-1)} \otimes 
    [1,1]^T
    \right) \otimes 
    [0,1]^T.
\end{align*}
Lemma \ref{lem:LR=RL} shows that the DRS algorithm splits $v^{(i+1)}$ and $(v^{(i+1)})'$ into the same number of columns.
Furthermore, Lemma \ref{lem:DRSforTwoVectors} shows that the number of output vectors of the DRS algorithm for 
  $  v^{(n)} = [v^{(i+1)}]^{(s_{i+2}, \ldots, s_{n})} $
and 
     $ (v^{(n)})' = [ (v^{(i+1)})' ]^{(s_{i+2}, \ldots, s_{n})}  $
are equal. 
    
Therefore, an occurrence of $(s_i, s_{i+1} )= (+,-)$ in a sequence can be replaced by $(s_i, s_{i+1} )= (-,+)$ without changing the number of output vectors of the DRS algorithm. 
Since any sequence $(s_1, s_2, \ldots, s_{n} )$  with $n_1$ minus signs and $n_2 $ plus signs can be permuted into ${(s'_1, s'_2, \ldots s'_{n})}, $ where $s'_i= -$ for $i\leq n_1$ and $s'_i= +$ for $i> n_1$, by repeatedly replacing any occurrence of $(+,-)$ by $(-,+)$, the above arguments show 
$n_{DRS}(v^{(n)}) =n_{DRS}(v^{(s'_1, s'_2, \ldots s'_{n})})$ always holds.
Hence, the size of DRS algorithm output for $v^{(n)}$ depends only on the values $n_1$ and $n_2$.  
\endproof

\vspace{2mm}

\subsubsection{ }\label{pf:prop:DRSrateloss}
\textit{Proof for Proposition \ref{prop:DRSrateloss}:}
First note that there is a bijection between $\mathset{-,+}^n$ and the columns of $G_2^{\otimes n}$ as follows.  
For each $\mathbf{s}= (s_1, \ldots, s_n)\in \mathset{-,+}^n$, there is exactly one column of $G_2^{\otimes n}$ in the form $1^{(\mathbf{s})}$ (see equation \eqref{eq:vi_recursion} and the paragraph following it, where we use $v= v^{(0)} =[1] \in \Ftwo$ to be the length-$1$ vector).
The term $\gamma$ can be characterized as follows:
\[
\gamma = \Big[ \frac{1}{N}\sum_{\mathbf{s}\in \mathset{-,+}^n} n_{DRS}(1^{(\mathbf{s})}) \Big] -1.
\]
By Lemma \ref{lem:rateLossIndepOfOrder}, the terms in the summation can be grouped according to the number of minus and plus signs in the sequence. Hence, 
\[
\gamma = \Big[\frac{1}{N}\sum_{i=0}^n \binom{n}{i} n_{DRS}(1^{(s_1= -, \ldots, s_i= -, s_{i+1}= +, \ldots, s_n= + ) } ) \Big] -1.
\]
Let $u_i$ denote the vector $1^{(s_1, \ldots, s_n)}$ with $s_l =-$ for $l\leq i$ and $s_l= +$ for $l> i$. 
Without loss of generality, let $n\lambda \in \N $.
For $i\leq n\lambda$, the Hamming weight of $u_i$ is $2^i \leq 2^{n\lambda} = w_{u.b.}$. Hence, $n_{DRS}(u_i) =1$. 
For $i > n\lambda$,  $u_i$ is split into $2^{i-n\lambda}$ vectors, each of which having weight equal to $2^{n\lambda}$. Therefore, $n_{DRS}(u_i)= 2^{i-n\lambda}$. 

The term $\gamma$ can be written as follows:
\begin{align}
    \gamma &= \sum_{i=0}^{n\lambda} \frac{1}{N}\binom{n}{i} +  \sum_{i=n\lambda+1}^n \frac{1}{N}\binom{n}{i}2^{i-n\lambda} -1  
    = \sum_{i=n\lambda+1}^n a_i , \label{eq:DRS_R_ai}
    \; 
\end{align} where $ a_i \triangleq \frac{1}{N}\binom{n}{i}(2^{i-n\lambda}-1)$.
Now, let $\alpha= i/n$. Since $i > n\lambda$ for each summand $a_i$, we consider $\alpha >\lambda >\frac{1}{2}$ in the following calculations. 
The term $a_i$ can be written as 
\begin{align}
    a_i = a_{n\alpha} = 2^{-n}\binom{n}{n\alpha} 2^{n\alpha-n\lambda +o(1)} 
    =  2^{-n}\cdot 2^{n h_b(\alpha) +o(1)} \cdot 2^{n\alpha-n\lambda +o(1)} 
    = 2^{n\cdot f(\alpha, \lambda) +o(1)}, \label{eq:DRS_ai_f} \; 
\end{align}
where the third equality is due to an asymptotic approximation of the binomial coefficient, and $f(\alpha,\lambda) \triangleq h_b(\alpha)+\alpha-\lambda -1$. 

 Consider $f(\alpha,\lambda)$ as a function of $\alpha$ over the interval $[0,1]$. We find its first and second derivatives with respect to $\alpha$ as follows:
\begin{align}\label{eq:fDeri}
     \frac{\partial f(\alpha,\lambda)}{\partial \alpha} = 1-\log{\frac{\alpha}{1-\alpha} }, \mbox{ and }
     \frac{\partial^2 f(\alpha,\lambda)}{\partial^2 \alpha} = 
     -\frac{1}{\ln{2}}\big( \frac{1}{ \alpha}+ \frac{1}{ 1-\alpha} \big) <0, 
     \mbox{ for any $0< \alpha <1$.}
\end{align}  
 Thus, for any fixed $\lambda$, $f(\alpha,\lambda)$ is a concave function of $\alpha$ and has local maximum when 
 $\frac{\partial f(\alpha,\lambda)}{\partial \alpha} =0 . $
From \eqref{eq:fDeri}, the equality holds if and only if  $\alpha= \frac{2}{3}$, and the maximum is 
 \begin{equation}
 \sup_{\alpha\in (0,1)}f(\alpha,\lambda) =
 	h_b\left(\frac{2}{3}\right)+\frac{2}{3}-\lambda -1
 = \lambda^*- \lambda .
 \label{eq:max_f_overalpha}
 \end{equation}
Also, when $\alpha =0$, $f(0,\lambda) = -\lambda -1$, and when  $\alpha =1$, $f(1,\lambda) = -\lambda$.  
Hence, $\sup_{\alpha\in [0,1]}f(\alpha,\lambda) = \lambda^*- \lambda.$
When $\lambda> \lambda^* $, we know $f(\frac{i}{n} ,\lambda) \leq \sup_{\alpha}f(\alpha,\lambda)  <0 $ for all integers $0 \leq i \leq n$, and  \eqref{eq:DRS_ai_f} implies that  $a_i \to 0$ exponentially fast for each $i$. 
 Equation \eqref{eq:DRS_R_ai} then shows that $\gamma$  vanishes exponentially fast in $n$.
\endproof

\subsection{Proof for Subsection \ref{subsection:LCdec_BEC}}\label{appendix:proof_LCdec_BEC}

\textit{Proof for Lemma \ref{lem:DRS_Z_BEC}:} 
We show the claim by proving the following:
when we encode the source bits according to $G'$, the bit-channels observed by the source bits are BECs and that the erasure probabilities are less than or equal to those when $G$ is used.  
We use proof by induction on  $n$.
For ease of notation, we use $M'$ to denote \textproc{DRS}$(M)$ for a given matrix $M$ in this proof.

\begin{figure}[h]
    \centering
    \includegraphics[width=0.25\textwidth]{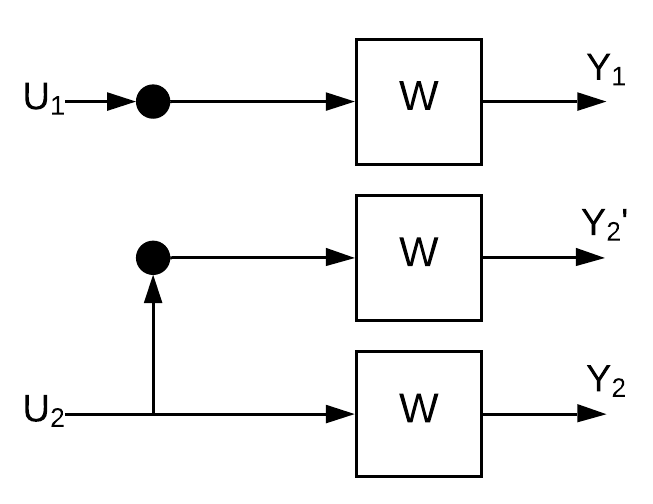}
    \caption{Encoding Block for $G_2'$}\label{fig:SplitG2}
    \vspace{-10mm}
\end{figure}

For $n =1$, if $G_2= G_2'$, we naturally have $Z_{G_2'}^{(s_1)} = Z_{G_2}^{(s_1)}$ for $s_1\in \mathset{-,+}$.
If $G_2 \neq G_2'$, the latter must be
${\tiny \begin{bmatrix}
		1 & 0 &0 \\
		0 & 1 & 1\\
\end{bmatrix}}$, corresponding to the encoding block diagram in Figure\,\ref{fig:SplitG2}, where the solid black circles indicate a \textit{split} of the XOR operation, i.e., the two operands of the original XOR operation are transmitted through two copies of channel $W$. 
The bit-channels observed by $U_1$ and $U_2$, denoted by $W^\boxminus$ and $W^\boxplus$,  are BECs with erasure probability $\epsilon$ and $\epsilon^2$, respectively. Note that the Bhattacharyya parameters satisfy the following: 
\begin{align*}
    Z_{G_2'}^{(-)}&= Z(W^\boxminus) =  \epsilon  \leq Z_{G_2}^{(-)} = Z(W^-) = 2\epsilon - \epsilon^2,  \\
    Z_{G_2'}^{(+)}&= Z(W^\boxplus) = \epsilon^2 = Z_{G_2}^{(+)} = Z(W^+).  
\end{align*}

Suppose that, for a fixed $w_{u.b.}$, the claim holds for all $n\leq m$ for some integer $m\geq 1$.
Let $B_m$ denote the encoding block corresponding to the generator matrix $(G_2^{\otimes m})'$, with inputs $U_1, \ldots, U_{2^m}$ and encoded bits  $X_1, \ldots, X_{f(m)}$, where $f(m)$ is the number of columns in $(G_2^{\otimes m})'$. 
Using the matrix notation, the relation between the 
input bits and encoded bits is:
\begin{equation*}\label{eq:WmMatrixNotation}
    (U_1, \ldots, U_{2^m}) (G_2^{\otimes m})' = (X_1, \ldots, X_{f(m)}). 
\end{equation*}
    
\begin{figure}
    \centering
    \includegraphics[trim=2cm 2.4cm 5.8cm 2cm, clip,  width=0.6 \textwidth]{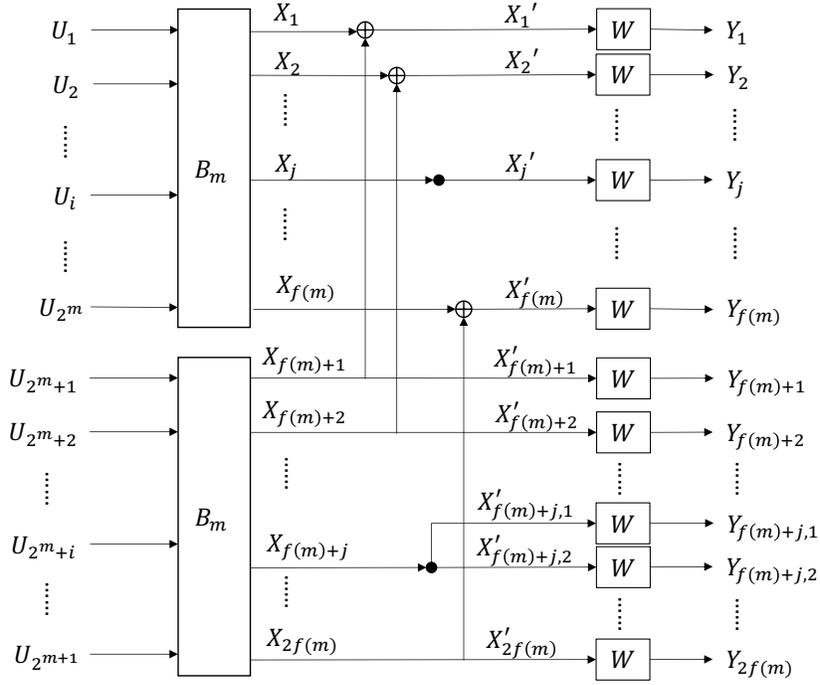}
    \caption{Encoding Block $B_{m+1}$}
    \label{fig:BmBlock}\vspace{-10mm}
\end{figure}

For $n= m+1$, the matrix $(G_2^{\otimes m+1})'$ is associated with $(G_2^{\otimes m})'$ as follows:
\begin{equation}\label{eq:DRS_G_recursion}
    (G_2^{\otimes m+1})' = \textproc{DRS}\Big(
\begin{bmatrix}
		(G_2^{\otimes m})' & \zero \\
		(G_2^{\otimes m})' & (G_2^{\otimes m})'
\end{bmatrix}\Big).
\end{equation}
Since $(G_2^{\otimes m})'$ consists of the outputs of the DRS algorithm, the columns in the right half of the input matrix in equation \eqref{eq:DRS_G_recursion} remain unaltered in the output. 
For the columns in the left half, they are of the form $[v^T, v^T]^T$ for some column $v$ of $(G_2^{\otimes m})'$.
If $2 w_H(v) > w_{u.b.}$, the outputs of the DRS algorithm are $[\zero^T, v^T]^T$ and $[v^T, \zero^T]^T$ because the vector $v$ must have weight no larger than the threshold. 
If  $2 w_H(v) \leq w_{u.b.}$, the algorithm leaves the vector unchanged.
We may represent the encoding block $B_{m+1}$ as in Figure\,\ref{fig:BmBlock}, where it is assumed that the $j$-th column of the input matrix in \eqref{eq:DRS_G_recursion} is halved by the DRS algorithm.

\begin{figure}
    \centering
    \includegraphics[trim=2cm 2.3cm 2.3cm 2cm, clip, width=0.7 \textwidth]{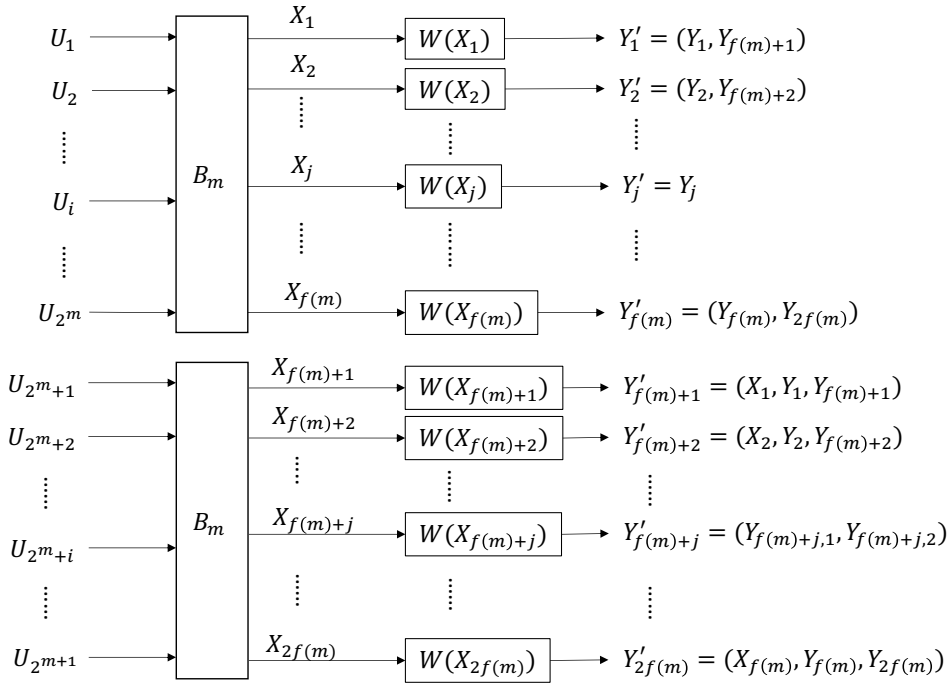}
    \caption{Equivalent Encoding Block $B_{m+1}$}
    \label{fig:BmBlock_Equvi}\vspace{-10mm}
\end{figure}

The erasure probabilities for the bit-channels observed by $X_i$, denoted here as $W(X_i)$, are less than or equal to $2\epsilon-\epsilon^2$ for $1\leq i \leq f(m)$, and are equal to $\epsilon^2$ for $f(m)+1 \leq i \leq 2 f(m)$, respectively.
Hence we may replace the XOR operations to the right of the $X_i$'s as well as the transmission over $W$'s by BECs $W(X_1), \ldots,W(X_{f{(m)}}),$ $W(X_{f{(m)}+1}), \ldots, W(X_{2f{(m)}})$, as in Figure\,\ref{fig:BmBlock_Equvi}.

One may observe that the erasure probability for the bit-channel observed by $U_i$ is a non-decreasing function of those of $W(X_1), \ldots, W(X_{f{(m)}})$ for  $i \leq 2^m$, and of $W(X_{f{(m)}+1}), \ldots,$  $W(X_{2f{(m)}})$ for  $i >2^m$. 
So for $i \leq 2^m$, we have  
\begin{align*}
    Z_{(G_2^{\otimes m+1})'}\left(U_i \,| \, Z(W)= \epsilon \right) 
    &\leq Z\left[ W(U_i)\, |\,  Z(W(X_j))= 2\epsilon- \epsilon^2, \mbox{ for } 1\leq j \leq f(m) \right]  \\
    = Z_{(G_2^{\otimes m})'} \left(U_{i} \,|\, Z(W)= 2\epsilon- \epsilon^2 \right) 
    &\leq Z_{G_2^{\otimes m}} \left(U_{i}\, | \, Z(W)= 2\epsilon- \epsilon^2 \right) 
    = Z_{G_2^{\otimes m+1}} \left(U_i \,| \,  Z(W)= \epsilon \right), 
\end{align*}
    where the first inequality is due to $Z(W(X_j))\leq  2\epsilon- \epsilon^2  \mbox{ for } 1\leq j \leq f(m)$ and the second inequality follows from the hypothesis of the induction. 
    
    Similarly, for  $i >2^m$, we have 
\begin{align*}
    Z_{(G_2^{\otimes m+1})'}\left(U_i  \,| \, Z(W)= \epsilon  \right) 
    &= Z\left[ W(U_i) | Z(W(X_j))=  \epsilon^2 ,\mbox{ for } f(m)+1\leq j \leq 2f(m)  \right]    \\
    = Z_{(G_2^{\otimes m})'} \left(U_{i-f(m)}  \,| \, Z(W)= \epsilon^2  \right)
    &\leq Z_{G_2^{\otimes m}} \left(U_{i-f(m)}  \,| \, Z(W)= \epsilon^2  \right) 
    = Z_{G_2^{\otimes m+1}} \left(U_i  \,| \, Z(W)= \epsilon  \right). 
\end{align*}  
Hence the inequality holds when $n= m+1$ as well. \endproof

\subsection{Proofs for Subsection \ref{subsection:LCdec_BMS}}
\label{appendix:proof_LCdec_BMS}

\subsubsection{ }\label{pf:prop:ADRScompl}
\textit{Proof of Proposition \ref{prop:ADRScompl}:}
First note that each XOR operation at the $j$-th recursion can be associated with exactly one vector $\mathbf{s} =(s_1, s_2, \ldots,  s_n) \in \mathset{-,+}^n$ and $s_j = -$. 
For example, assume that the number of minus signs in $\mathbf{s}$, denoted as $m(\mathbf{s})$, is larger than $n_{lub} =\log w_{u.b.} = n\lambda$, and let $\tau =\tau(\mathbf{s})$ be the index such that $m(s_{\tau}, s_{\tau+1},\ldots, s_n)= n_{lub} $ and $s_{\tau} =-$. 
Then for each index $i$ in the set $\mathset{k: 1 \leq k < \tau, s_k =-}$, there is a bijection between the pair $(\mathbf{s}, i)$ and an XOR operation at the $i$-recursion which is split and modified in the ADRS scheme. 
Hence, the extra complexity of the SC decoder for the ADRS scheme, compared to that of the SC decoder for the code based on $G_2^{\otimes n}$, is given by
\begin{align}\label{eq:ADRScomp_bound}
    \sum_{l=1}^{n-n_{lub}+1} & \abs{\mathset{\mathbf{s} \in \mathset{-,+}^n: \tau(\mathbf{s} ) > l , s_l =- } }  2(2^{l+1} -2)c \nonumber \\
    = &\sum_{l=1}^{n-n_{lub}+1} \sum_{k=l+1}^{n-n_{lub}+1} \abs{\mathset{\mathbf{s} \in \mathset{-,+}^n: \tau(\mathbf{s} ) =k , s_l =-  } }  2(2^{l+1} -2)c \nonumber \\
    =  &\sum_{k=1}^{n-n_{lub}+1} \left[ \binom{n-k+1}{n_{lub} } 2^{k-2} \sum_{l=1}^{k-1}    2(2^{l+1} -2)c \right] \nonumber 
    \leq  4c\sum_{k=1}^{n-n_{lub}+1} \binom{n-k+1}{ n_{lub}} 2^{k-2}  \sum_{l=1}^{k-1}   2^{l}\nonumber  \\
    =\, &4c\sum_{k=1}^{n-n_{lub}+1} \binom{n-k+1}{ n_{lub}}  2^{k-2}  (2^{k}-2)  
    \leq  4c\sum_{k=0}^{n-n_{lub}} \binom{n-k}{ n_{lub}}  2^{2k}.
\end{align}
Now, let  $\alpha = \frac{k}{n} \in [0, 1-\lambda]$. Using Stirling's approximation we have
\begin{align*}
    &\binom{n-k}{ n_{lub}}  2^{2k} =  \binom{n(1-\alpha)}{ n\lambda }  2^{2\alpha n} \approx 2^{n(1-\alpha) h_b(\frac{\lambda }{1-\alpha}) +2\alpha n},
    \end{align*} for all sufficiently large $n$.
It suffices to assume that $\lambda < \frac{3}{4}$. For $\lambda \geq \frac34$, note that fewer XOR operations are split and modified, and that the resulting additional decoding complexity is not larger than when $\lambda < \frac34$ is used. 
Now let $f(\alpha, \lambda) = (1-\alpha) h_b(\frac{\lambda}{1-\alpha}) +2\alpha$.
We find its maximum, for a given $\lambda$, by solving 
\begin{align*}
0 &=\frac{\partial}{\partial \alpha}f(\alpha, \lambda) =\frac{\partial}{\partial \alpha}\left[ -(1-\alpha) (\frac{\lambda}{1-\alpha} \log \frac{\lambda}{1-\alpha})- 
(1-\alpha)(1-\frac{\lambda}{1-\alpha}) \log (1-\frac{\lambda}{1-\alpha}) +2\alpha
\right] \\
&=\frac{1}{\ln 2} \left[
-\ln(1-\alpha)+ \ln(1-\alpha -\lambda)
\right] +2,
\end{align*}
which is true if and only if $\alpha = 1- \frac{4}{3}\lambda$.
And note that \[
\frac{\partial^2}{\partial \alpha^2}f(\alpha, \lambda) = \frac{1}{\ln 2}\left[ \frac{1}{1-\alpha} - \frac{1}{1-\alpha -\lambda} \right] <0 
\] for all $\alpha  \in [0, 1-\lambda]$.
The maximum of the function is then given by 
$f(1- \frac{4}{3}\lambda, \lambda) = 2 -  \lambda \log3.$
Using the union bound, the sum in \eqref{eq:ADRScomp_bound} can be bounded by $
4c n 2^{n( 2 - \lambda \log 3 )}
$, and the ratio of the sum to  $N= 2^n$, denoted by $\gamma_C$, is bounded from above as
$
\gamma_C \leq 4c n 2^{n( 1 -\lambda \log 3 )}.
$
Since the exponent $n( 1 - \lambda \log 3)$ goes to negative infinity as $n$ grows when $\lambda > \lambda^{\dagger} = 1/\log3 \approx 0.631$, the ratio $\gamma_C$ vanishes exponentially in $n$ when  $\lambda >\lambda^{\dagger}$. 
The proposition follows by noting that the SC decoding complexity for the code based on $G_2^{\otimes n}$ is $N\log N$.
\endproof

\vspace{2mm}

\subsubsection{ }\label{pf:Prop:BMSrateloss}
\textit{Proof of Proposition\,\ref{Prop:BMSrateloss}:}
Similar to the proof of Proposition\,\ref{prop:ADRScompl}, the number of additional channels due to the ADRS scheme modification is given by 
\begin{align}\label{eq:ADRSrate_bound}
    \sum_{l=1}^{n-n_{lub}+1}  \abs{\mathset{\mathbf{s} \in \mathset{-,+}^n: \tau(\mathbf{s} ) > l , s_l =- } }  2^l \nonumber 
    = &\sum_{l=1}^{n-n_{lub}+1} \sum_{k=l+1}^{n-n_{lub}+1} \abs{\mathset{\mathbf{s} \in \mathset{-,+}^n: \tau(\mathbf{s} ) =k , s_l =-  } }  2^l \nonumber \\
    \leq  \sum_{k=1}^{n-n_{lub}+1} \binom{n-k+1}{ n_{lub}} 2^{k-2}  \sum_{l=1}^{k-1}   2^{l}  
    \leq \, &\sum_{k=0}^{n-n_{lub}} \binom{n-k}{ n_{lub}}  2^{2k}.
\end{align}
By the argument in the proof of Proposition\,\ref{prop:ADRScompl}, the sum in \eqref{eq:ADRSrate_bound} can be upper bounded by 
$ n 2^{n( 2 - \lambda \log 3 )}$, and the ratio of the sum to  $N= 2^n$, denoted by $\gamma$, is bounded from above as
$ \gamma  \leq  n 2^{n( 1 -\lambda \log 3 )}.$
Since the exponent $n( 1 - \lambda \log 3)$ goes to negative infinity as $n$ grows when $\lambda > \lambda^{\dagger} = 1/\log_2 3 \approx 0.631$, the ratio $\gamma $ vanishes exponentially in $n$ when  $\lambda >\lambda^{\dagger}$. 
\endproof

\vspace{3mm}

\bibliographystyle{IEEEtran}
\bibliography{IEEEabrv}

\end{document}